\documentclass[prb,twocolumn,showpacs,superscriptaddress]{revtex4-1}

\usepackage[english]{babel}
\usepackage[applemac]{inputenc}
\usepackage[T1]{fontenc}
\usepackage{ae}
\usepackage{graphicx} 
\usepackage{amsfonts} 
\usepackage{amssymb} 
\usepackage{amsmath} 
\usepackage{latexsym} 
\usepackage{psfrag} 
\usepackage{float} 
\usepackage{color} 
\usepackage[raggedright, nooneline]{subfigure} 
\subfiguretopcaptrue
\usepackage{natbib}

\newcommand{\ket}[1]{\mid\! #1\rangle}
\newcommand{\bra}[1]{\langle #1\!\mid}

\newcommand{\beq}{\begin{equation}}
\newcommand{\eeq}{\end{equation}}

\newcommand{\Rea}{{\Re \text{e}}}

\newcommand{\bk}{\mathbf{k}}
\newcommand{\ad}{a^{\phantom{\dagger}}}
\newcommand{\ac}{a^\dagger}
\newcommand{\fk}{f^{\phantom{\star}}_k}
\newcommand{\fks}{f^\star_k}
\newcommand{\Ai}{\mathrm{Ai}}
\newcommand{\Bi}{\mathrm{Bi}}
\newcommand{\bw}{\begin{widetext}}
\newcommand{\ew}{\end{widetext}}

\begin{document}

\title{Dynamics following  a linear ramps  in the $O(N)$ model: dynamical transition and statistics of excitations}


\author{Anna Maraga}
\affiliation{SISSA, International School for Advanced Studies, via Bonomea 265, 34136 Trieste, Italy}
\author{Pietro Smacchia}
\affiliation{Department of Physics and Astronomy, Rutgers University, Piscataway, New Jersey 08854, USA}
\author{Alessandro Silva}
\affiliation{SISSA, International School for Advanced Studies, via Bonomea 265, 34136 Trieste, Italy}
\affiliation{Abdus Salam ICTP, Strada Costiera 11, 34100 Trieste, Italy}

\date{\today} 

\pacs{...}

\begin{abstract}
Non-thermal dynamical critical behavior can arise in isolated quantum systems brought out of equilibirum by a change in time of their parameters. While this phenomenon has been studied in a variety of systems in the case of a sudden quench, here we consider its sensitivity to a change of protocol by considering the experimentally relevant case of a linear ramp in time. Focusing on the $O(N)$ model in the large $N$ limit, we show that a dynamical phase transition is always present for all ramp durations and discuss the resulting crossover between the sudden quench transition and one dominated by the equilibrium quantum critical point. We show that the critical behavior of the statistics of the excitations, signaling the non-thermal nature of the transition are robust against changing protocol. An intriguing crossover in the equal time correlation function, related to an anomalous coarsening is also discussed.
\end{abstract}

\maketitle


\section{Introduction}
\label{sec:Introduction}

The nonequilibrium dynamics of isolated quantum many-body systems has been the subject of many theoretical and experimental studies~\cite{Polkovnikov2011, Lamacraft2012, Dziarmaga2010, yukalov2011} in recent years.
The interest in this field is mainly motivated by the advances in the experimental study of cold atoms trapped in optical lattices~\cite{bloch_review}.
These systems are characterized by a very weak coupling to the external environment, which strongly suppresses dissipative and decoherence effects and allows the observation of the coherent quantum many-body dynamics for quite long time scales.
In this context, a series of remarkable experiments led, for example, to the observation of the collapse and revival of a system driven across the Mott-superfluid transition~\cite{greiner2002a, greiner2002b}, the spontaneous symmetry breaking in a quenched spinor Bose-Einstein condensate~\cite{sadler_06}, the absence of thermalization in a one-dimensional Bose gas~\cite{kinoshita}, the phenomenon of prethermalization~\cite{Kitagawa2011, Gring2012, langen2013}, and the light-cone spreading of correlations~\cite{cheneau_2012}.

Among all the possible ways of taking an isolated quantum system out of equilibrium, the most natural one is to vary in time one of its parameters. 
A natural goal of any experimental and theoretical characterizations of  nonequilibrium dynamics is to be able to predict the nature of the steady state attained by a system long after such variation has occurred.
While generic systems are expected to approach a thermal state~\cite{deutsch_91, srednicki_94, rigol_08} even when thermally isolated from the environment, in special cases
(i.e. for integrable systems~\cite{kinoshita, rigol_07, barthel_08, kollar_08, iucci_09, cazalilla2012}) relaxation to a nonthermal state described by the Generalized Gibbs Ensemble (GGE)~\cite{Jaynes1957} consistent with all the constants of motion is anticipated.
Despite the peculiar nature of integrable systems, signatures of non-thermal behaviour may be observed even in non-integrable ones: the relaxation to a thermal state may indeed involve the approach to a nonthermal quasistationary state (prethermal state)~\cite{Berges2004, moeckel_08, moeckel_09, moeckel_10, Kollar2011, Marino2012, Marcuzzi2013, Essler2014}  on intermediate time scales. Such prethermal states are either expected in low dimensions for systems approximately integrable, as well as in the presence of long range interactions and in large dimensions close to a mean field limit. Most importantly, recent literature has shown that such 
quasistationary states may display dynamical critical behaviour. 
Originally studied for sudden changes of parameters (quenches) in the Hubbard model~\cite{eckstein_2009, schiro_2010, Tsuji2013} such criticality was later observed in several systems at the mean-field level~\cite{sciolla_2010, sciolla_2011} and in field theories~\cite{Gambassi2011a, sciolla_2013, Smacchia2015, Chiocchetta2015}.
While the characterization of these dynamical transitions and their peculiarity as compared to thermal transition is a topic of recent research, it has been recently shown that
a simple protocol measuring the statistics of excitations produced in a sudden quench can single out their non-equilibrium nature ~\cite{Smacchia2015}.

In general, any dynamical evolution is expected to depend on the particular protocol selected to vary the system parameters. While dynamical transitions were studied for
instantaneous variations (sudden quenches) considering more generic procedures, such as a linear ramp, could shed some light on which dynamical features are unaffected by the changes of the protocol and which ones depend on its details (for example, on its duration).
Moreover, the study of generic protocols can be useful for eventual experiments, which typically use linear ramps to prepare and study particular states. In this work we 
therefore address the sensitivity of dynamical transitions to a change of protocol, from a sudden quench to a linear ramp. 
We focus on the case of an $O(N)$ vector model
in the large $N$ limit, where the model can be solved exactly~\cite{moshe_2003}, driving the system out of equilibrium by a linear variation in time of the bare mass, starting in the  disordered phase.
We will show that for this system the dynamical phase transition is robust against changing the protocol and map entirely the crossover
between a true dynamical transition and one dominated by the equilibrium quantum critical point as a function of the duration of the ramp $\tau$.
We will in particular discuss analytically the location of the dynamical critical point $r_c$ as a function of ramp duration, focusing in particular on the
two limits of large and small $\tau$. While both critical exponents as well as the behavior of the statistics of excitations in a double quench are found 
to be hardly sensitive on the change of protocol, we observe an intriguing crossover  in the equal time correlation functions displaying anomalous coarsening.

The paper is organized as follows. 
In Sec.~\ref{sec:Model} we review the critical properties of the system at equilibrium and in the case of a sudden quench in the bare mass.
In Sec.~\ref{sec:Ramp-Dynamics} we study the dynamics of the system when a linear ramp is performed, detecting the dynamical critical point and computing the critical dimensions and exponents.
The characterization of the dynamical transition based on the statistics of excitations is discussed in Sec.~\ref{sec:Ramp-Excitations}, while the case of a linear ramp below the dynamical critical point is studied in Sec.~\ref{sec:Ramp-below}. 
In Sec.~\ref{sec:Conlusions} we summarize the results.

 
\section{The model}
\label{sec:Model}

In the following we will focus on the dynamical phase transition of an interacting $N$ component real scalar field $\vec{\phi}$ in $d$ spatial dimensions, described by the Hamiltonian 
\beq \label{eq:H}
\mathcal{H}= \frac{1}{2} \! \int \! d^d x \left[ \big( \vec{\Pi} \big)^2 \! + \big( \nabla \vec{\phi} \big)^2 \! + r_0 \big( \vec{\phi} \big)^2 \! + \frac{\lambda}{12 N} \big( \vec{\phi} \big)^4 \right] ,
\eeq 
where $\vec{\Pi}$ is the conjugate momentum field. We will be interested in characterizing the dynamical phase transition occurring in the mean field,  $N \to \infty$ limit, where the $O(N)$ vector model is exactly solvable~\cite{moshe_2003}.
In this limit  and at equilibrium this system is described by a quadratic theory with an effective mass $r$, satisfying the self-consistent equation
\beq\label{sconst}
r=r_0+\frac{\lambda}{6}\int dr\;\langle  \phi^2 \rangle, 
\eeq
where exploiting the O(N) symmetry of the model we focused on one of the components of the field $\vec{\phi}$, indicated as $\phi$. 
Using this equation one may easily see that the system exhibits both a quantum and a thermal phase transition between a paramagnetic phase and an ordered one, 
characterized by the spontaneous breaking of the $O(N)$ symmetry~\cite{moshe_2003}.
At the critical point, identified by the vanishing of the effective mass~$r$, the bare mass is given by
\beq \label{eq:rc_equilibrium}
r_0^c= -\frac{\lambda}{12} \int^{\Lambda} \!\!\! \frac{d^d k}{(2 \pi)^d} \, \frac{1}{k} \coth\!\left( \frac{\beta k}{2} \right),
\eeq
where $\Lambda$ is the ultraviolet cutoff and $\beta$ is the inverse temperature. 
The integral on the right hand side converges for $d>2$ ($d>1$ at zero temperature), setting therefore the value for the lower critical dimension.
Moreover, one can compute the critical exponent $\nu$ describing the divergent behavior of the correlation length~$\xi \sim r^{-1}$ close to the critical point, i.e., $\xi \sim (\delta r _0)^{-\nu}$, with $\delta r _0= r_0 - r_0^c$. 
At $T=0$, one finds $\nu=1/(d-1)$ for $1<d<3$, and $\nu=1/2$ for $d\geq3$, which is therefore the upper critical dimension of the quantum phase transition. 
In the finite temperature case, one gets instead $\nu=1/(d-2)$ for $2<d<4$, and $\nu=1/2$ for $d\geq4$, which implies that $d=4$ is the upper critical dimension for the thermal transition.



Focusing now on the dynamics, it has been shown numerically~\cite{SC10, sciolla_2013, Chandran2013, Smacchia2015} that this model can undergo a dynamical phase transition after a sudden quench in the bare mass, i.e., suddenly changing its value from $r_{0, i}$ to $r_{0,f}$ (
we focus here on the case of a quench starting from the ground state in the paramagnetic phase). The time dependent effective mass  (satisfying Eq.(\ref{sconst}) with a time dependent correlation function $\langle \phi^2(t) \rangle $ dictating the self-consistency) is seen to oscillate and then relax to a well defined value at large times. The stationary value $r^{\star}$ of the effective mass can be predicted efficiently via an ansatz~\cite{SC10} (see below) based on the replacement of the equal time correlation function 
$\langle \phi^2(t)\rangle$ in
Eq.(\ref{sconst}) with the stationary, time averaged part of corresponding post-quench correlator for a free theory ($\lambda=0$) with  the initial and final values of the mass 
set equal to $r_i$ and $r^{\star}$. 
The dynamical critical point is therefore reached provided the final bare mass satisfies the relation
\beq \label{eq:rc_quench}
r_{0,f}^c=-\frac{\lambda}{24} \int^{\Lambda} \!\!\! \frac{d^d k}{(2 \pi)^d} \frac{2 k^2 + r_i}{k^2\sqrt{k^2 + r_i}},
\eeq
where $r_i$ indicates the effective mass before the quench.

From this equation one obtains that 
the lower critical dimension for the dynamical transition is $d=2$, and that the dynamical critical point is always smaller than the quantum critical point $r_0^c$.
As in the equilibrium case, we denote with $\xi^*$ the correlation length in the stationary state and with $\nu^*$ the exponent describing its divergence close to the dynamical critical point. We find that $\nu^*=1/(d-2)$ for $2<d<4$, and $\nu^*=1/2$ for $d\geq4$, which is the upper critical dimension.
The fact that these critical exponents are similar to those of a thermal transition at equilibrium suggest that the two might be analogous~\cite{Chandran2013,Smacchia2015}. 
Indeed, one could imagine that fixing $r_0^f$ in the equilibrium ordered phase and increasing $r_0^i$ from $r_0^f$ to higher values amounts to increase the energy density injected by the quench into the system. This could be seen as equivalent to moving from low to high temperatures in the corresponding equilibrium phase diagram, in which case a 
thermal phase transition would sooner or later be crossed. Notice however that, despite the analogies, the distribution of quasi-particles after a quench  in the $N\rightarrow +\infty$ limit \it is not \rm thermal. Moreover, the difference between the two cases becomes apparent if one studies the statistics of excitations produced close to a dynamical transition, since, unlike the equilibrium case, in the dynamical one the fluctuations in the number of excitations are very sensitive to how close one is to a dynamical critical point~\cite{Smacchia2015}.  


\section{Dynamics and dynamical critical properties for a linear ramp}
\label{sec:Ramp-Dynamics}

In this paper we address the robustness of the scenario above with respect to a change of protocol from a sudden quench to a linear ramp
of the bare mass. 
The system is initially prepared in the ground state of the disordered phase ($r_{0,i}>r_0^c$), then the bare mass is linearly decreased to a final value $r_{0,f}$, according to the following protocol: $r_0(t)=r_{0,i}$ for $t < 0$, $r_0(t)=r_{0,i}+ (r_{0,f}-r_{0,i}) t/\tau$ for $0 \leq t \leq \tau$, and $r_0(t)= r_{0,f}$ for $t > \tau$.

Let us start setting up the formalism to study the dynamics in the $N\rightarrow +\infty$ limit.
The system is again described by an effective quadratic Hamiltonian with a time dependent effective mass $r(t)$. Exploiting the $O(N)$ symmetry of the model, we can focus on only one component of the field. 
Passing to Fourier space  we may write
\beq
\mathcal{H}_\mathrm{eff}(t)=\frac{1}{2} \int^{\Lambda} \!\!\! \frac{d^d k}{(2 \pi)^d} \left[ \Pi_\bk(t) \Pi_{-\bk}(t) + \omega_k^2(t) \phi_\bk(t) \phi_{-\bk}(t) \right], \label{eq:H_eff} 
\eeq
where  $\omega_k(t)=\sqrt{k^2+r(t)}$, and 
\beq
r(t)=r_0(t) + \frac{\lambda}{6} \int^{\Lambda} \!\!\! \frac{d^d k}{(2 \pi)^d} \langle \phi_\bk(t) \phi_{-\bk}(t)\rangle. \label{eq:rt1}
\eeq
Let us now expand the field in the Heisenberg representation as
\beq \label{eq:phi}
\phi_\bk(t) = \fk (t) \ad_\bk + \fks (t) \ac_{-\bk},
\eeq
where $\ad_\bk$ and $\ac_\bk$ diagonalize the initial Hamiltonian~(\ref{eq:H_eff}) at $t=0$ and $ \fk (t)$ is a complex amplitude.
Imposing the Heisenberg equations of motion for $\phi_\bk(t)$, we derive the equation for the evolution of the mode function $\fk (t)$
\begin{subequations} \label{eq:f-r}
\beq \label{eq:f}
\ddot{f}^{\phantom{\star}}_k (t) +\left[ k^2+r(t) \right] \fk (t) = 0,
\eeq
where
\beq \label{eq:rt2}
r(t) =r_0(t) + \frac{\lambda}{6} \int^{\Lambda} \!\!\! \frac{d^d k}{(2 \pi)^d} \lvert \fk (t) \rvert^2
\eeq
\end{subequations}
and the initial conditions are $\fk (0)=1/\sqrt{2 \omega_k(0)}$ and $\dot{f}^{\phantom{\star}}_k(0)=-i\sqrt{\omega_k(0)/2}$, with $\omega_k(0)=\sqrt{k^2+r_i}$. 

\begin{figure}[t]
\centering%
\includegraphics[width=0.85\columnwidth]{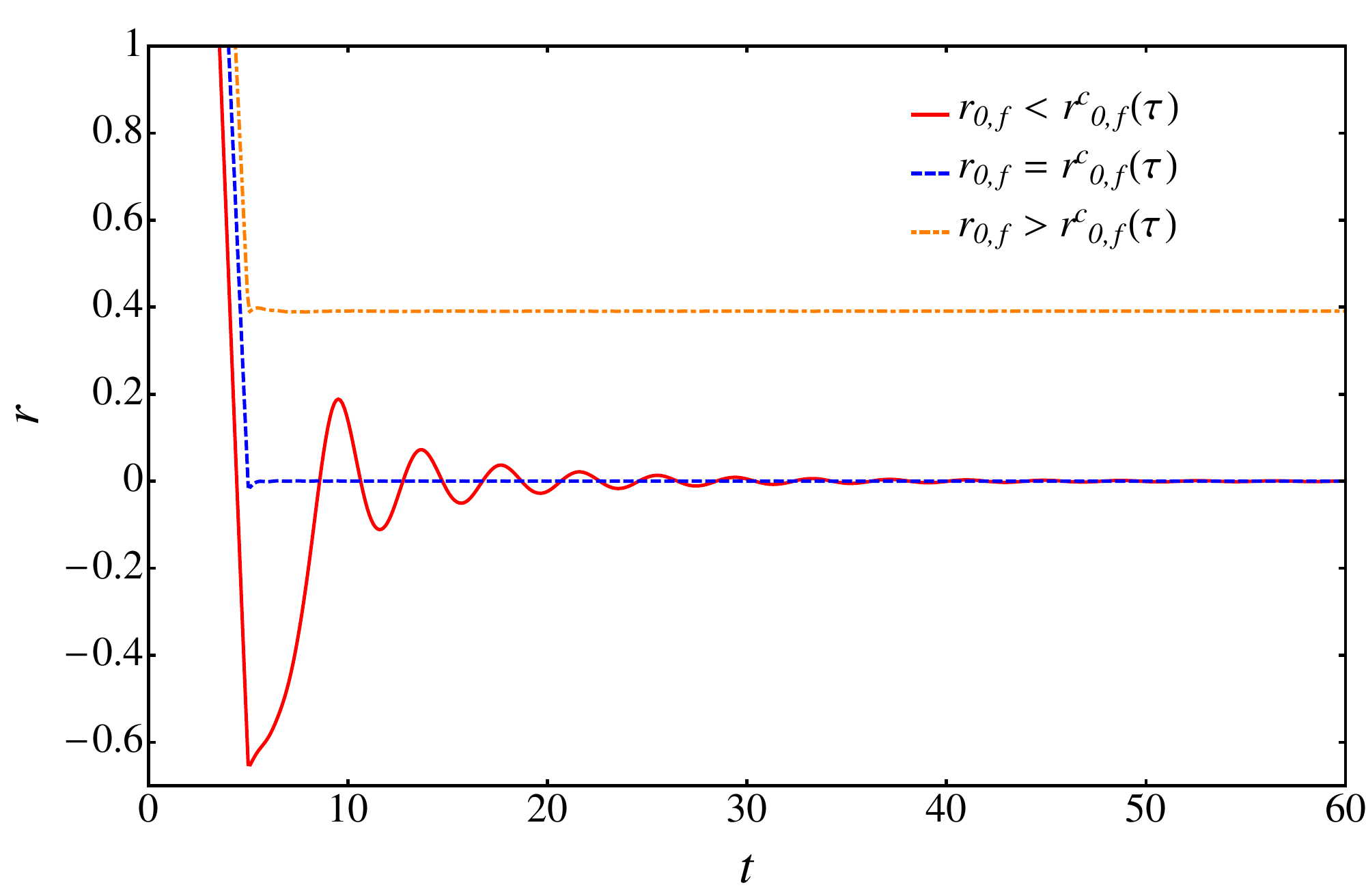}
\caption{(Color online) Time evolution of the effective mass~$r(t)$ for ramps of duration~$\tau = 5$, initial bare mass~$r_{0,i}=5$ in~$d=3$ and interaction strength $\lambda=15$. Final values of the bare mass below, at, and above the dynamical critical point are shown.}%
\label{fig:r}
\end{figure}

These equations have the same form as those obtained for a quench, with the only difference that $r_0$ is now not a constant but a linear function of time.
In particular, Eq.~(\ref{eq:f-r}) can be solved analytically for a linear ramp in the special case of $\lambda=0$ (see Appendix~\ref{app:A}).
For any finite $\lambda$ one instead has to resort to numerical integration. 
Varying the duration of the ramp and the value of the final bare mass, the system is found to display again a dynamical phase transition: as shown in Fig.~\ref{fig:r}, long after the end of the ramp the effective mass $r(t)$ is seen to relax to a stationary value, which is positive up to a certain $\tau$-dependent dynamical critical point $r_{0,f}^c(\tau)$, and vanishes for $r_{0,f}\le r_{0,f}^c(\tau)$.

\subsection{Stationary state and dynamical criticality}\label{statstate}

Let us now  characterize thoroughly the dynamical phase transition as a function of initial and final parameters and ramp duration $\tau$. It is first of all important to be able, as in the case of a sudden quench, to predict analytically the stationary value of the effective mass $r^*$. 
In order to achieve this goal, we  introduced an ansatz for the stationary effective mass inspired by the one used before for a sudden quench~\cite{SC10, Smacchia2015}: we assume the stationary part of the equal time Green's function $\langle \phi_\bk(t) \phi_{-\bk}(t)\rangle=\lvert \fk(t) \rvert^2$ to be equal to the non-interacting ($\lambda=0$) one, with the bare masses replaced by the renormalized ones, namely $r_{0,i}\rightarrow r_i$ (which can be calculated with Eq.(\ref{eq:f-r}) with $t<0$) and $r_{0,f}\rightarrow r^*$ (see Appendix~\ref{app:A}).
We therefore obtain the following self-consistent equation for $r^*$
\beq \label{eq:r*_ramp}
r^* \! =r_{0,f}+ \frac{\lambda}{12} \int^{\Lambda} \!\!\! \frac{d^d k}{(2 \pi)^d} \! \left[ \lvert f^0_k (r^*\!, \tilde{\tau}) \rvert^2 + \frac{\lvert \dot{f}^0_k (r^*\!, \tilde{\tau}) \rvert^2}{k^2 + r^*} \right],
\eeq
where $f^0_k$ denotes the mode function for $\lambda=0$ (see Eq.(\ref{mode1})-(\ref{mode2})).
According to this ansatz, we can identify the dynamical critical point at which $r^*$ vanishes as
\beq \label{eq:rc_ramp}
r_{0,f}^c(\tau)=- \frac{\lambda}{12} \int^{\Lambda} \!\!\! \frac{d^d k}{(2 \pi)^d} \! \left[ \lvert f^0_k (0, \tilde{\tau}) \rvert^2 + \frac{\lvert \dot{f}^0_k (0,\tilde{\tau}) \rvert^2}{k^2} \right].
\eeq
The mere fact that the stationary state can be described by an ansatz such as Eq.~(\ref{eq:r*_ramp}) allows to deduce many of the properties of the dynamical phase transition. Note, however, that in order to obtain the correct stationary value for $r_{0,f} \geq r_{0,f}^c(\tau)$ we had to renormalize the ramp duration $\tau$ to an effective value $\tilde{\tau}$ in Eq.~(\ref{eq:r*_ramp}). Such renormalized value increases as $\tau$ does (see discussion below).

\begin{figure}[t]
\centering%
\includegraphics[width=0.85\columnwidth]{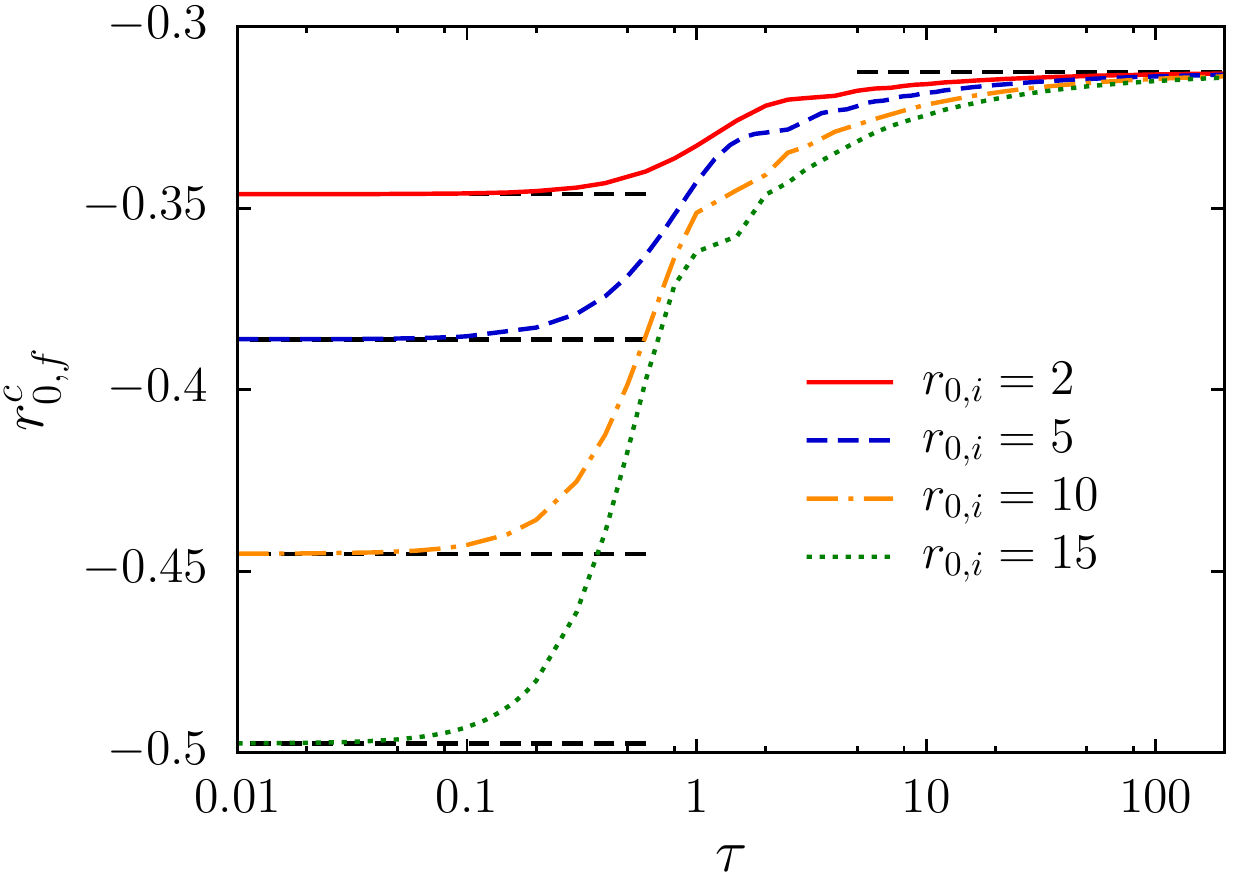}
\caption{(Color online) Dynamical critical point $r_{0,f}^c(\tau)$ as a function the ramp duration $\tau$ in $d=3$. Different values of the initial bare mass $r_{0,i}$ are shown, while the interaction is~$\lambda=15$. Horizontal black dashed lines indicate the dynamical critical point for a sudden quench ($\tau \to 0$) and the quantum critical point for the equilibrium case ($\tau \to \infty$).}%
\label{fig:r_critico}
\end{figure}

Let us now establish the lower critical dimension of the dynamical transition by analyzing the behavior for low momenta of the integrand of Eq.~(\ref{eq:rc_ramp}) (see Appendix~\ref{app:B}). Inspection of Eq.(\ref{eq:rc_ramp}) gives that
for every finite $\tau$, the modes that contribute the most to the integral on the right hand side are those with $k \ll  {\rm Min}[(r_i/\tilde{\tau})^{1/3}, \sqrt{r_i}]$, where both $\lvert f^0_k (0, \tilde{\tau}) \rvert^2 $ and $\lvert \dot{f}^0_k (0, \tilde{\tau}) \rvert^2$ go to a constant, making
 the integrand behave as $1/k^2$. This implies that the dynamical critical point~$r_{0,f}^c(\tau)$ is finite for $d > 2$, $d = 2$ being the lower critical dimension for every finite~$\tau$.
We observe that as $\tau$ increases, the region considered above shrinks. Moreover,   as $\tau$ gets larger and larger the region of intermediate asymptotics $(r_i/\tilde{\tau})^{1/3} \ll k \ll \sqrt{r_i}$, where $\lvert f^0_k (0, \tilde{\tau}) \rvert^2 \sim 1/k$ and $\lvert \dot{f}^0_k (0, \tilde{\tau}) \rvert^2 \sim k$, becomes more and more important.
When $\tau$ becomes infinite this asymptotics dominates and the lower critical dimension becomes $d = 1$, recovering the result of the quantum transition. As shown in Fig.~\ref{fig:r_critico}, the values of $r_{0,f}^c(\tau)$ interpolate between the dynamical critical point for a sudden quench, corresponding to $\tau \to 0$, and the quantum critical point at equilibrium, in the limit of large $\tau$.

\begin{figure}[t]
\centering%
\subfigure[]{\includegraphics[width=.52\columnwidth]{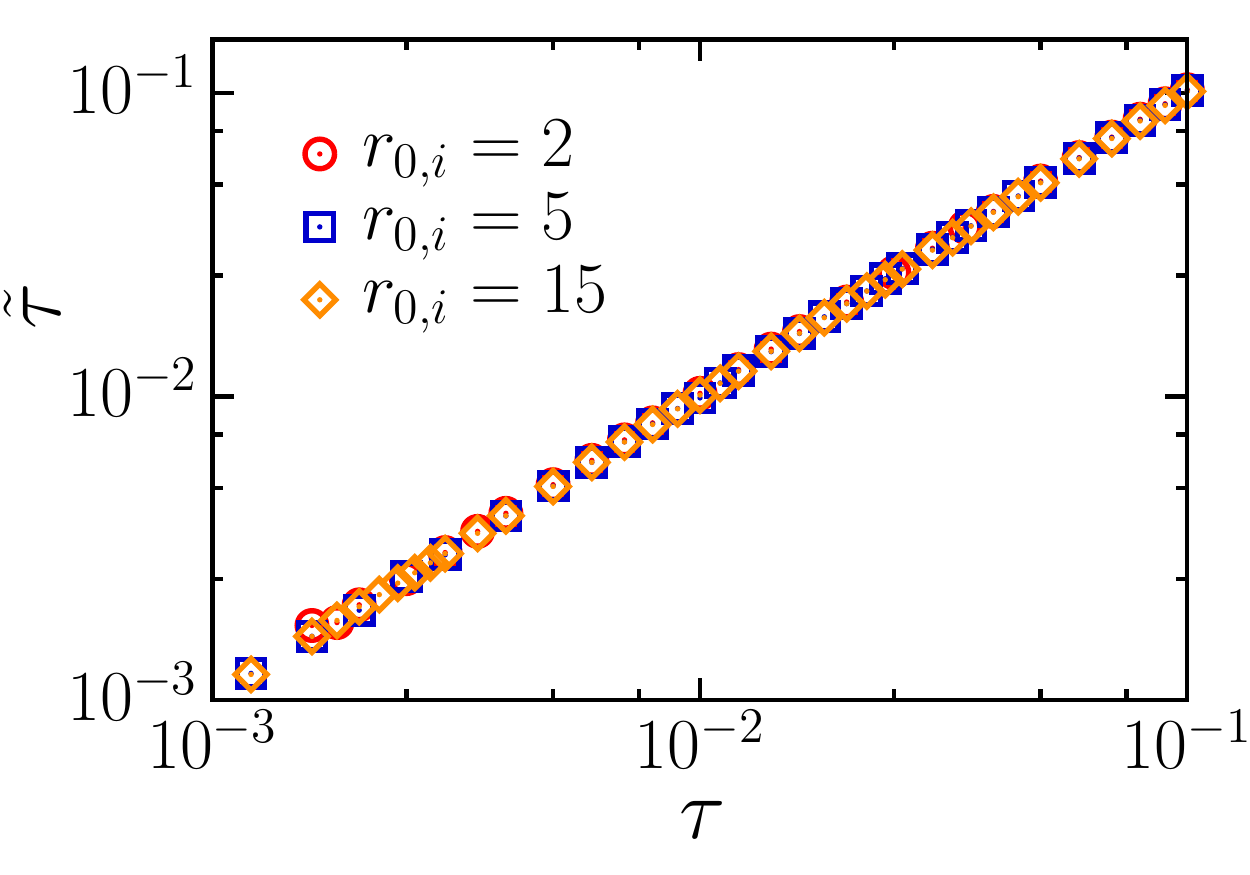}\label{fig:taueff_small3d}}%
\subfigure[]{\hspace{-0.27cm}\includegraphics[width=.52\columnwidth]{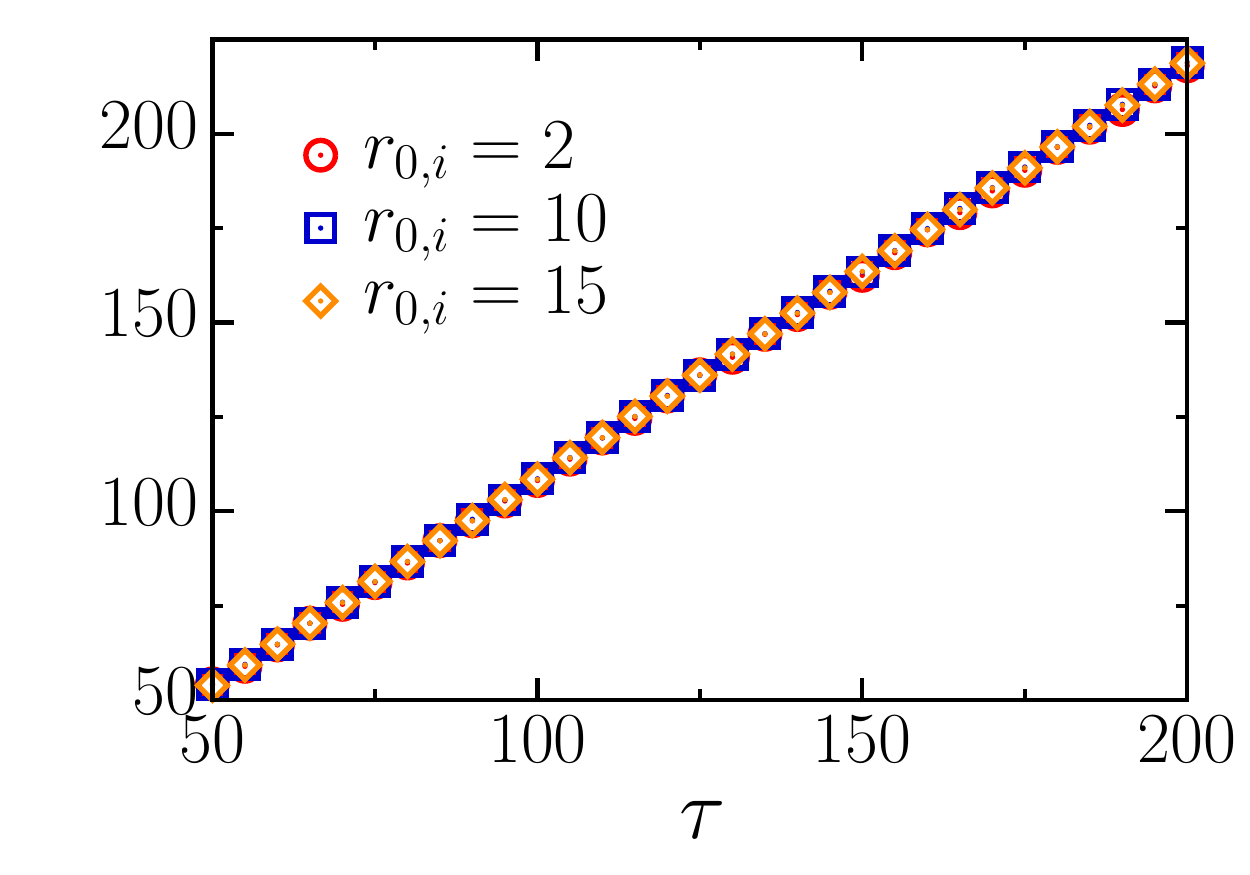}\label{fig:taueff_large3d}}
\caption{(Color online) Effective ramp duration $\tilde{\tau}$ as a function of the true ramp duration $\tau$ at the dynamical critical point in $d=3$ for small~\subref{fig:taueff_small3d} and large~\subref{fig:taueff_large3d} $\tau$. Different values of the initial bare mass $r_{0,i}$ are shown, while the interaction is $\lambda=15$.}%
\label{fig:taueff}
\end{figure}

It is now important to study the dependence of $\tilde{\tau}$ on $\tau$. 
Eq.~(\ref{eq:r*_ramp}) provides the correct stationary value of the effective mass provided the parameter $\tilde{\tau}$ is adjusted, a task that can be accomplished 
numerically. In particular, once the dynamical critical point has been identified, we can compute {\sl a posteriori} the effective ramp duration $\tilde{\tau}$ at criticality using Eq.~(\ref{eq:rc_ramp}).
Analyzing the behavior of $\tilde{\tau}$ as a function of the true ramp duration $\tau$ at the critical point and for $r_{0,i}$ and $\lambda$ fixed, it turns out that in the limits of small and large $\tau$ these two quantities have a linear relation, as can be seen in Fig.~\ref{fig:taueff}.
Moreover, varying the value of the initial bare mass $r_{0,i}$ (but keeping $\lambda$ fixed) the different $\tilde{\tau}(\tau)$ collapse on the same line, for large and small $\tau$. 
We may therefore use the ansatz~(\ref{eq:rc_ramp}) to  analytically study  how the dynamical critical value depends on $\tau$ in two limiting cases, for $\tau \to \infty$ (adiabatic switching) and $\tau \to 0$ (sudden quench).

Using the linear relation between $\tau$ and $\tilde{\tau}$ at large $\tau$ we may now employ Eq.~(\ref{eq:rc_ramp}) to study the crossover in Fig.(\ref{fig:r_critico}).
We will use in particular the exact solutions for the non-interacting mode functions $f^0_k(t)$ expressed in terms of Airy functions (see Appendix~\ref{app:A}). 
Employing the asymptotic expansion of the Airy functions for large and negative arguments (see Appendix~\ref{app:B}), for $\tilde{\tau} \gg 1/\sqrt{r_i}$ Eq.~(\ref{eq:rc_ramp}) reads
\beq \label{eq:rc_large}
r_{0,f}^c(\tau)\simeq- \frac{\lambda}{12} \frac{\Omega(d)}{(2 \pi)^d} \left( \mathcal{I}_1(d) + \mathcal{I}_2(d)  \right),
\eeq
where $\Omega(d)$ is the solid angle in $d$ dimension and
\bw
\begin{subequations} \label{eq:I}
\begin{align}
&\mathcal{I}_1(d)= \frac{\pi}{4}\, \Lambda^d  \left( \frac{\tilde{\tau}}{r_i} \right)^{1/3} \! \int_0^1 \!\! d z \, z^\frac{d-2}{2} \left[ \Ai^2 \left( - \frac{\Lambda^2 \tilde{\tau}^{2/3}}{r_i^{2/3}} \, z \right) + \Bi^2 \left(- \frac{\Lambda^2 \tilde{\tau}^{2/3}}{r_i^{2/3}} \, z \right) \right], \\
&\mathcal{I}_2(d)=\frac{\pi}{4}\, \Lambda^{d-2}  \left( \frac{r_i}{\tilde{\tau}} \right)^{1/3} \! \int_0^1 \!\! d z \, z^\frac{d-4}{2} \left[ \Ai'^{\,2} \left( - \frac{\Lambda^2 \tilde{\tau}^{2/3}}{r_i^{2/3}} \, z \right) + \Bi'^{\,2} \left(- \frac{\Lambda^2 \tilde{\tau}^{2/3}}{r_i^{2/3}} \, z \right) \right],
\end{align}
\end{subequations}
\ew
where we introduced the dimensionless variable $z=k^2/\Lambda^2$.

Integrals in Eq.~(\ref{eq:I}) can be computed exactly both in $d=3$ and $d=4$ (see Appendix~\ref{app:C}).
We find that the asymptotic value of the dynamical critical point for large~$\tau$ and $d=3$ is
\beq \label{eq:rc_large-3d} 
r_{0,f}^c(\tau) = r_0^c + \frac{\lambda \,\, \Gamma(-1/3)}{2^{17/3} \! \cdot \! 3^{7/3} \, \pi^2} \left(\frac{r_i}{\tilde{\tau}} \right)^{2/3} \! + O \! \left( \! \frac{r_i^{4/3}}{\Lambda^4 \tilde{\tau}^{4/3}} \! \right),
\eeq
while for $d=4$ is
\beq \label{eq:rc_large-4d}
r_{0,f}^c(\tau) = r_0^c - \frac{\lambda}{1152 \sqrt{3} \,\pi^2 }  \left(\frac{r_i}{\tilde{\tau}} \right)  + O \! \left( \! \frac{r_i^2}{\Lambda^6 \tilde{\tau}^2} \! \right),
\eeq
where $r_0^c$ is the quantum critical point at equilibrium (see Eq.~(\ref{eq:rc_equilibrium})). In both cases $r^c_{0,f}$ is smaller than the equilibrium  critical point.

\begin{figure}[t]
\centering%
\subfigure[]{\includegraphics[width=.52\columnwidth]{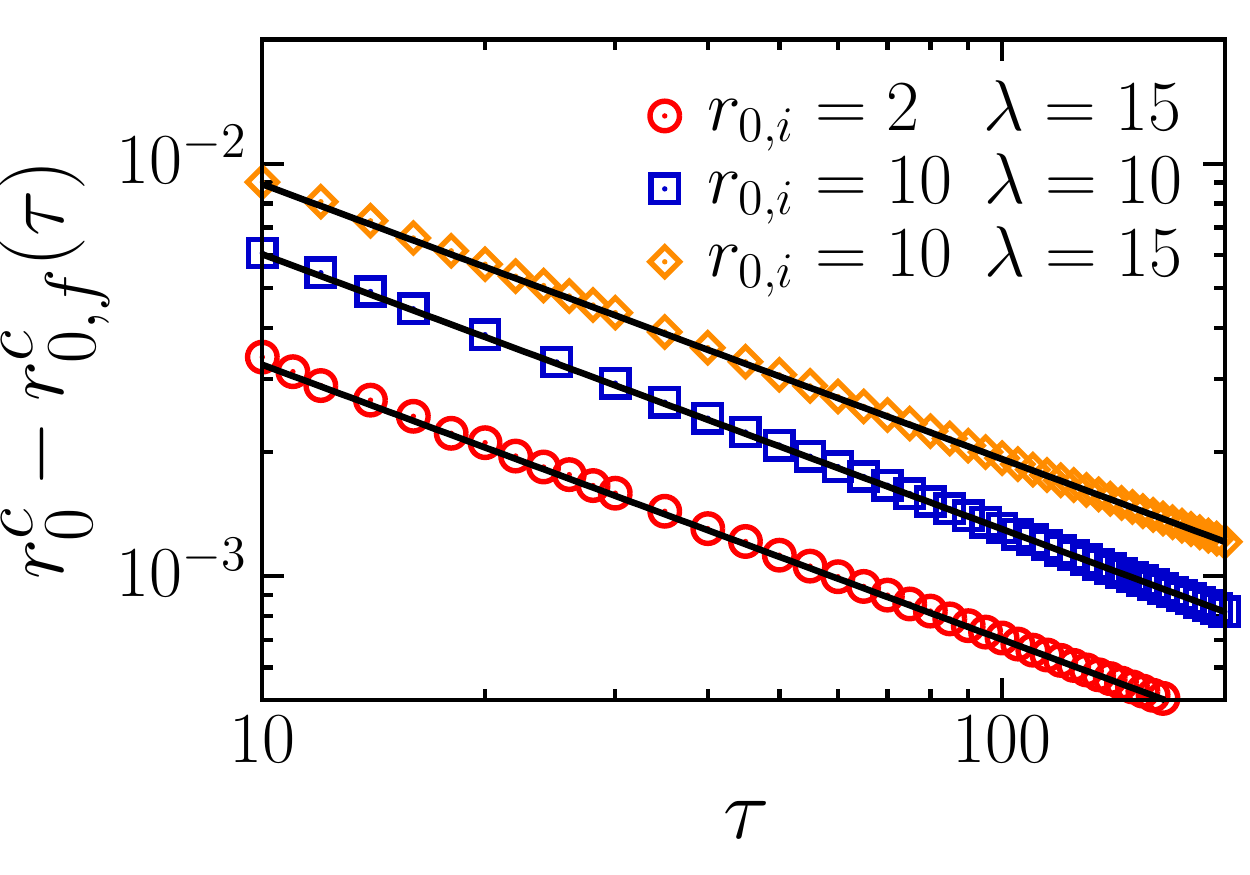}\label{fig:rc_large3d}}%
\subfigure[]{\hspace{-0.37cm} \includegraphics[width=.52\columnwidth]{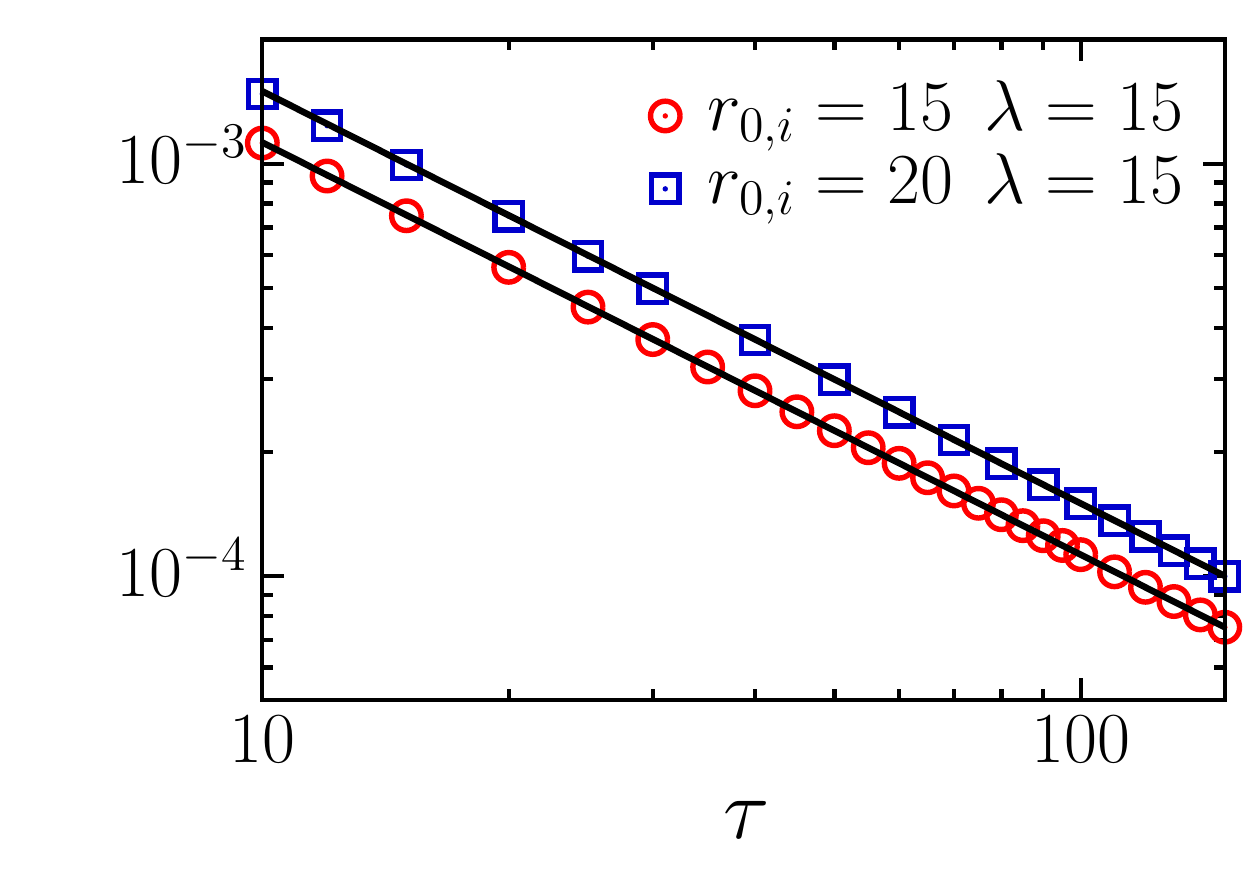}\label{fig:rc_large4d}}
\caption{(Color online) Difference between the quantum critical point $r_0^c$ and the dynamical critical point $r_{0,f}^c(\tau)$ for large ramp duration $\tau$ in $d=3$~\subref{fig:rc_large3d} and $d=4$~\subref{fig:rc_large4d}. Different values of the interaction $\lambda$ and of the initial bare mass $r_{0,i}$ are shown. Black lines are proportional to $\tau^{-2/3}$~\subref{fig:rc_large3d} and to $\tau^{-1}$~\subref{fig:rc_large4d}.}%
\label{fig:rc_large}
\end{figure}

Since for large $\tau$ the relation between $\tilde{\tau}$ and $\tau$ is linear at the critical point, we conclude that the dynamical critical point approaches the quantum critical value as~$\tau^{-2/3}$ for $d=3$ and as $\tau^{-1}$ for $d=4$.
We verified these scalings numerically by linearly fitting the relation between $\tilde{\tau}$ and $\tau$ for large $\tau$ and replacing the result in Eqs.~(\ref{eq:rc_large-3d}) and~(\ref{eq:rc_large-4d}), getting an excellent agreement with numerical data, as shown in Fig.~\ref{fig:rc_large}.

Let us now consider the fate of dynamical critical point in the limit of small $\tau$. 
By using the asymptotic expansion of the Airy functions for small arguments (see Appendix~\ref{app:B}), we have that
\begin{align}
&\lvert f^0_k (0, \tilde{\tau}) \rvert^2 \simeq \frac{1}{2 \sqrt{k^2+r_i}} + \frac{r_i}{6\sqrt{k^2+r_i}} \, \tilde{\tau}^2, \\
&\lvert \dot{f}^0_k (0, \tilde{\tau}) \rvert^2 \simeq  \frac{\sqrt{k^2+r_i}}{2} - \frac{4k^2 r_i+r_i^2}{24 \sqrt{k^2+r_i}} \,\tilde{\tau}^2.
\end{align}
Inserting these expressions in Eq.~(\ref{eq:rc_ramp}), we obtain
\beq
r_{0,f}^c(\tau)\simeq r_{0,f}^c(0) + \frac{\lambda}{12} \,\tilde{\tau}^2 \! \int^{\Lambda} \!\!\! \frac{d^d k}{(2 \pi)^d}\frac{r_i^2}{24 k^2\sqrt{k^2 + r_i}},
\eeq
where $r_{0,f}^c(0)$ is the dynamical critical point for a sudden quench (see Eq.~(\ref{eq:rc_quench})).

\begin{figure}[t]
\centering%
\subfigure[]{\includegraphics[width=.52\columnwidth]{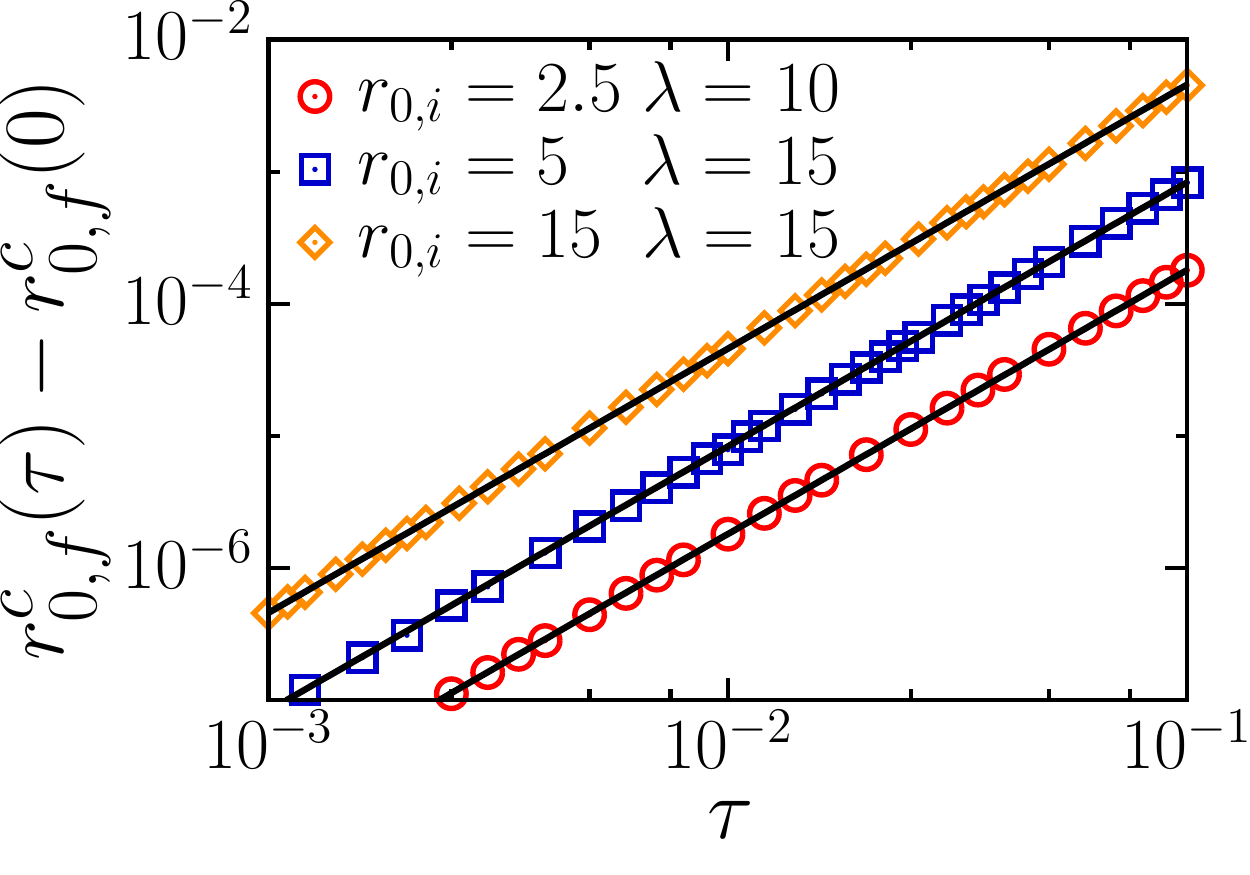}\label{fig:rc_small3d}}%
\subfigure[]{\hspace{-0.4cm} \includegraphics[width=.52\columnwidth]{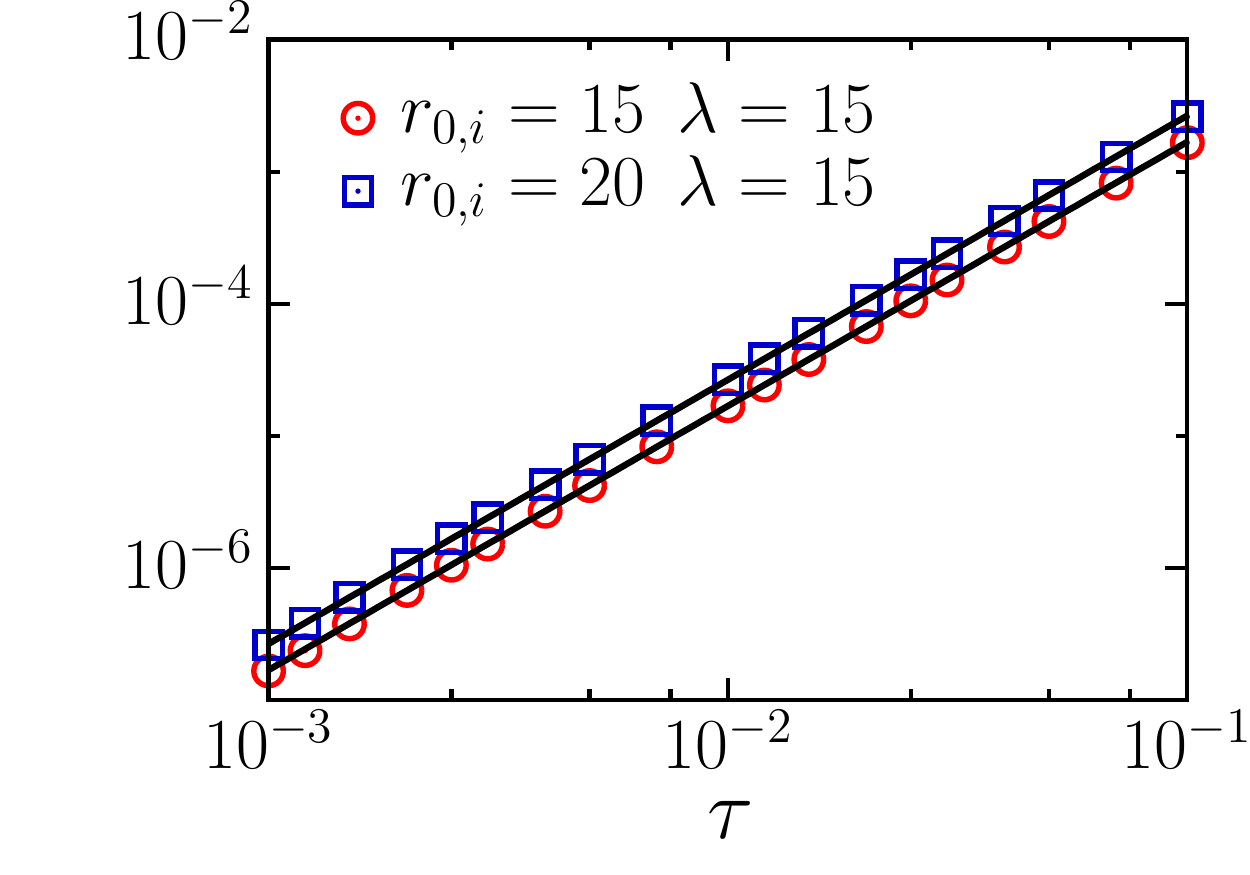}\label{fig:rc_small4d}}
\caption{(Color online) Difference between the dynamical critical point for a ramp, $r_{0,f}^c(\tau)$, and for a sudden quench, $r_{0,f}^c(0)$, for small ramp duration $\tau$ in $d=3$~\subref{fig:rc_small3d} and $d=4$~\subref{fig:rc_small4d}. Different values of the interaction $\lambda$ and of the initial bare mass $r_{0,i}$ are shown. Black lines are proportional to $\tau^2$.}%
\label{fig:rc_small}
\end{figure}

Since at criticality $\tilde{\tau} \sim \tau$ for small $\tau$, we conclude that the dynamical critical point departs from the sudden quench value as $\tau^2$, both in $d=3$ and $d=4$.
This is confirmed by numerical data (Fig.~\ref{fig:rc_small}).

\begin{figure}[t]
\centering%
\subfigure[]{\includegraphics[width=.51\columnwidth]{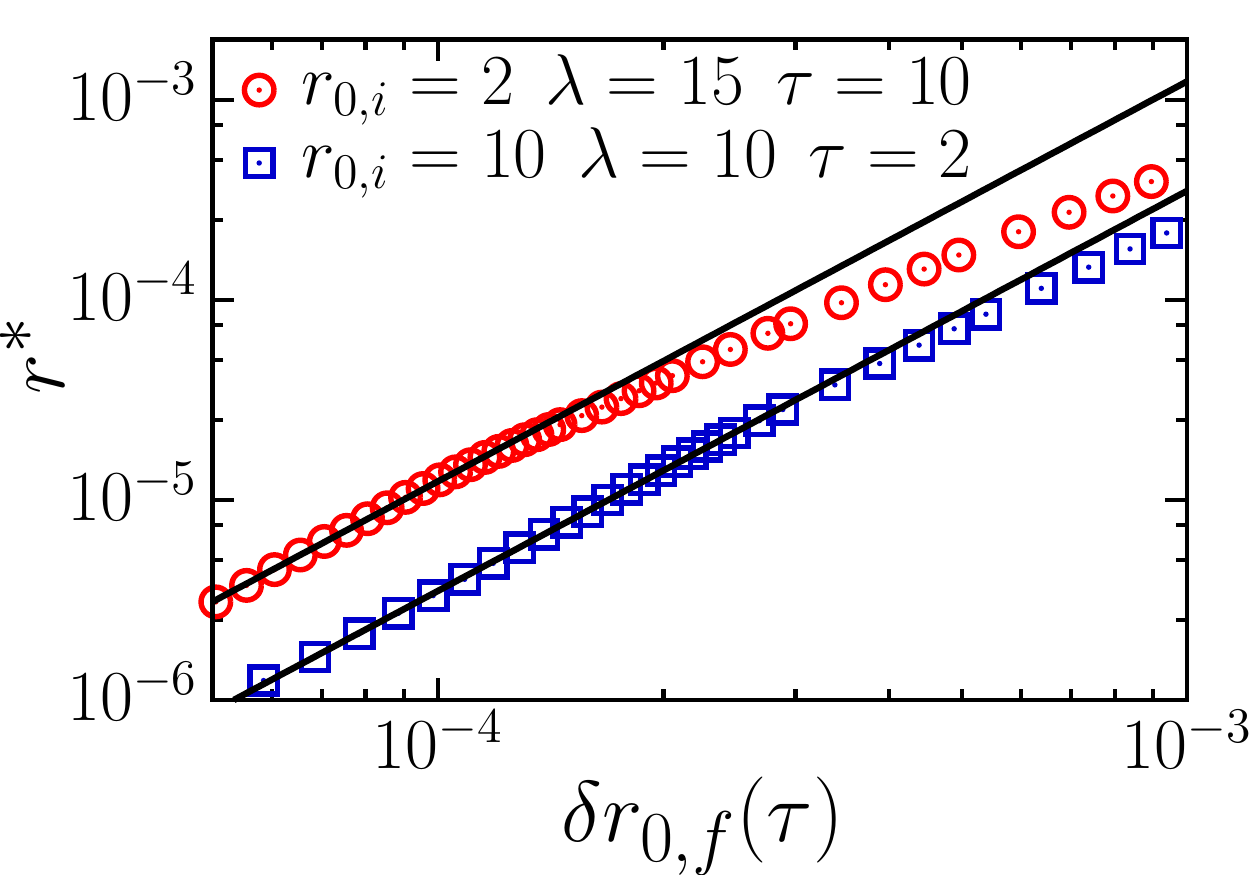}\label{fig:nu*3d}}%
\subfigure[]{\hspace{-0.1cm}\includegraphics[width=.51\columnwidth]{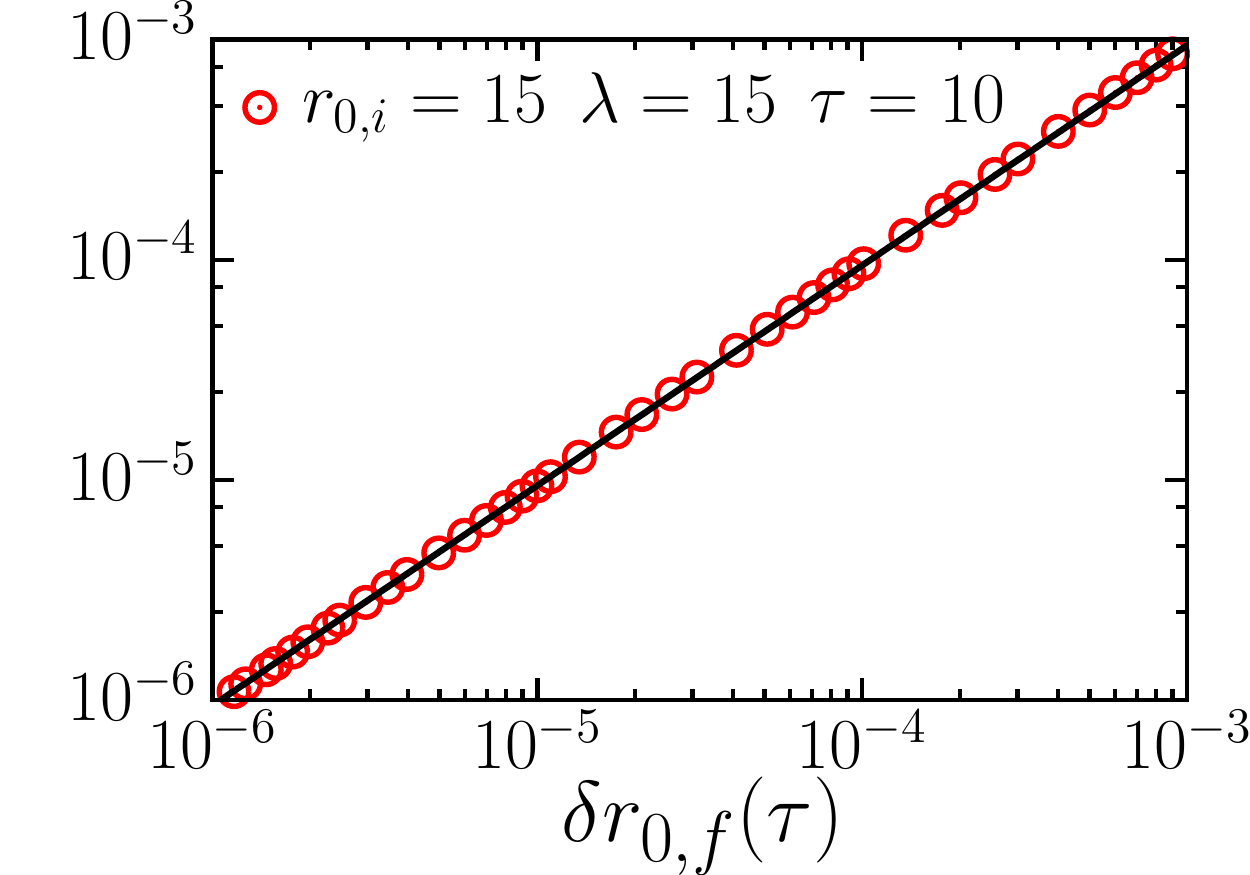}\label{fig:nu*4d}}
\caption{(Color online) Stationary value of the mass as a function of the distance from the dynamical critical point in~$d=3$~\subref{fig:nu*3d} and $d=4$~\subref{fig:nu*4d}. Black lines are quadratic~\subref{fig:nu*3d} and linear~\subref{fig:nu*4d} fits.}%
\label{fig:nu*}
\end{figure}

We are now ready to compute the critical exponent $\nu^*$, describing the divergence of the correlation length $\xi^*$ in the stationary state close to the dynamical critical point, i.e., $\xi^* \sim (\delta r_{0, f}(\tau))^{-\nu^*}$, with $\delta r _{0, f}(\tau)= r_{0,f} - r_{0,f}^c(\tau)$ combining Eqs.~(\ref{eq:r*_ramp}) and~(\ref{eq:rc_ramp}). 
As shown in detail in Appendix~\ref{app:B}, for $2<d<4$ the stationary value of effective mass at the leading order scales as $r^* \sim (\delta r_{0, f}(\tau))^\frac{2}{d-2}$, while for $d \ge 4$ the scaling becomes linear, i.e., $r^* \sim \delta r_{0, f}(\tau)$.
Since the theory is Gaussian, $(\xi^*)^{-1} \sim \sqrt{r^*}$. 
We conclude that
\beq \label{eq:nu*}
\begin{array}{lll}
\nu^*= 1/(d-2) & \text{ for} & 2<d<4, \\
\nu^*= 1/2 & \text{ for} & d \ge 4,
\end{array}
\eeq
$d = 4$ being the upper critical dimension.
Fig.~\ref{fig:nu*} shows that numerical results for $d=3$ and $d=4$ agree with this prediction.
We note that for $d=3$ (Fig.~\ref{fig:nu*3d}) numerical data follow the relation $r^* \sim (\delta r_{0, f}(\tau))^2$ for sufficiently small values of $r^*$ and then depart from this scaling, eventually approaching a linear relation for larger~$r^*$, indicating a crossover between $d=3$ critical and mean field behaviour.  

As in the case of a quench, the critical dimensions and the critical exponent turn out to be the same as the thermal one, even though 
we are dealing here with the unitary dynamics of a pure state and not with a mixed state. 
Only when $\tau$ is strictly infinite we eventually recover the results of the quantum transition.

\section{Statistics of excitations}
\label{sec:Ramp-Excitations}

\begin{figure*}[t]
\centering%
\subfigure[]{\includegraphics[width=.32\textwidth]{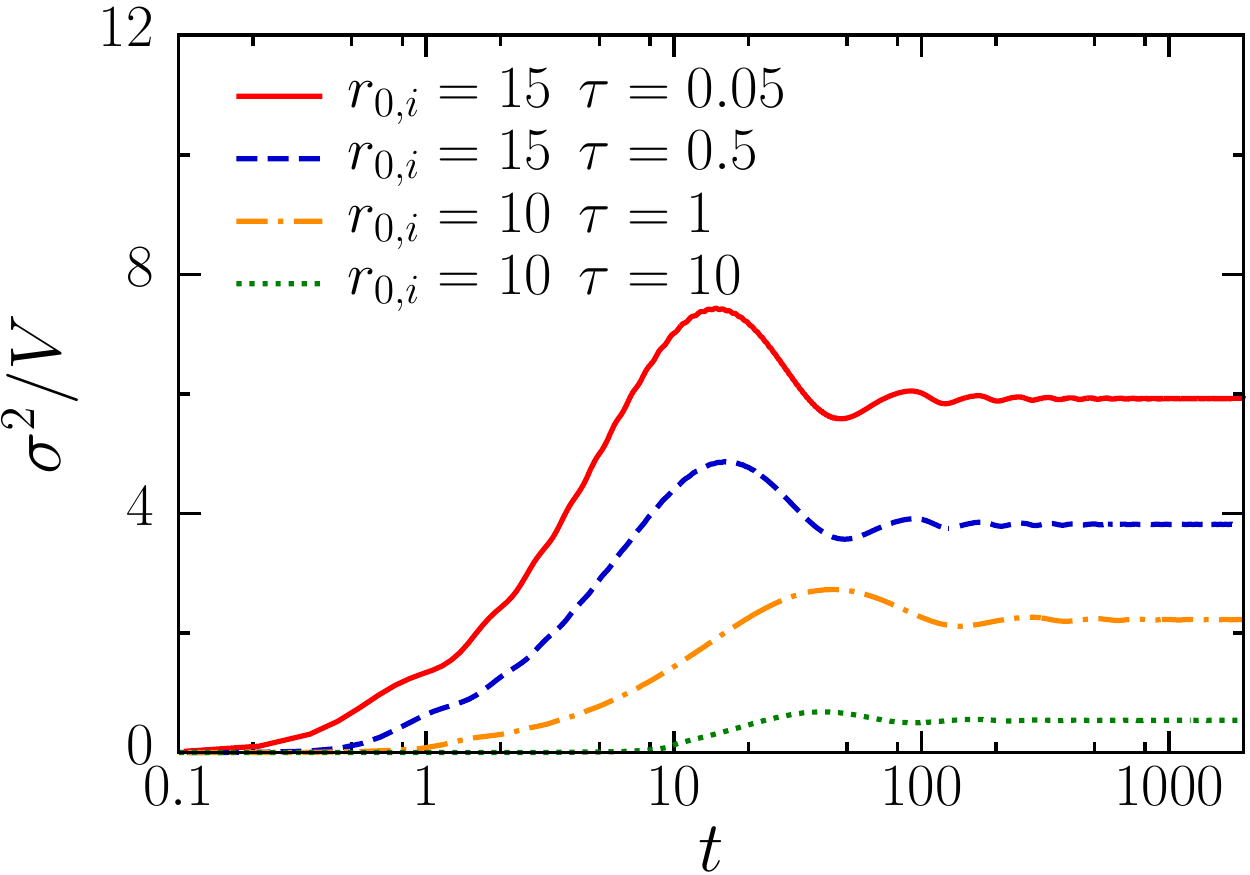}\label{fig:sigma3d>}}%
\subfigure[]{\hspace{0.27cm}\includegraphics[width=.32\textwidth]{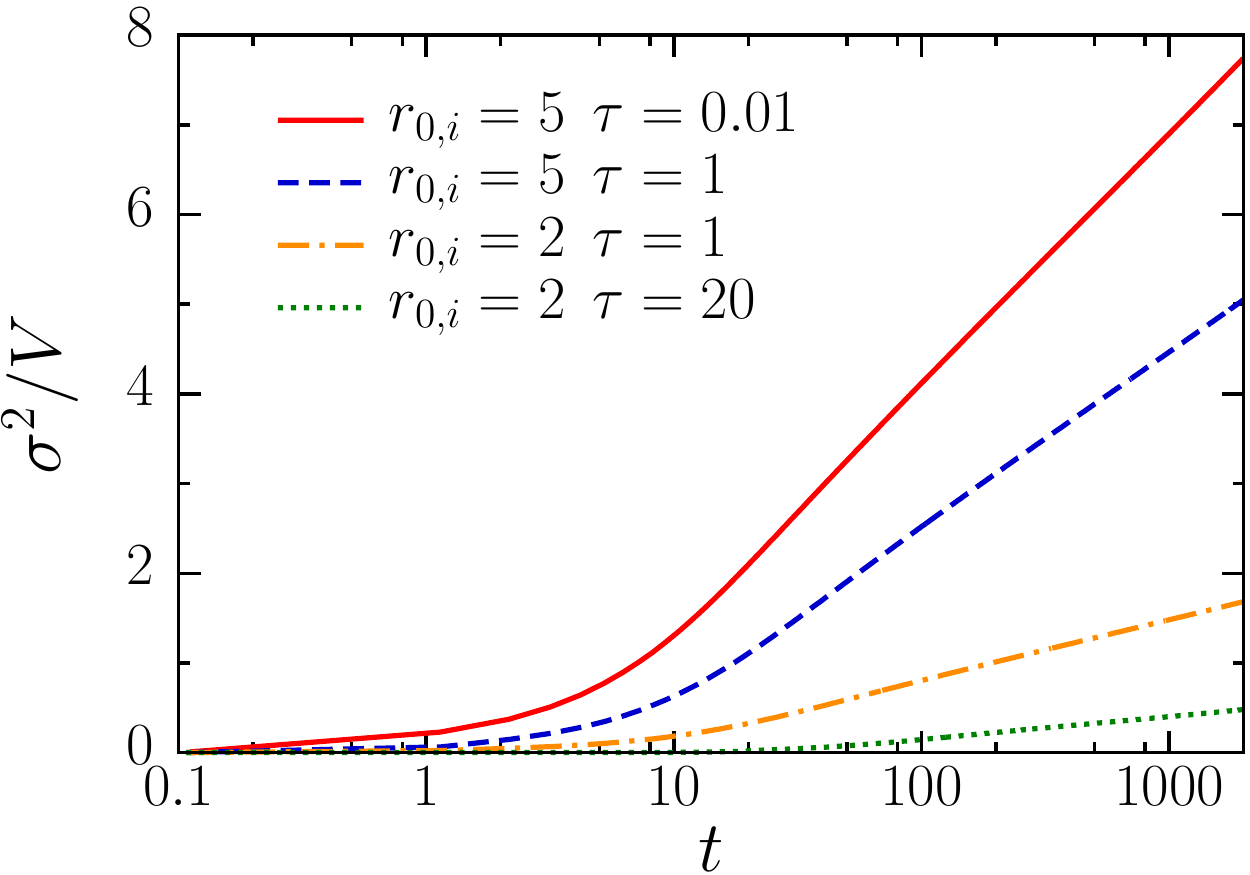}\label{fig:sigma3d=}}%
\subfigure[]{\hspace{0.27cm}\includegraphics[width=.32\textwidth]{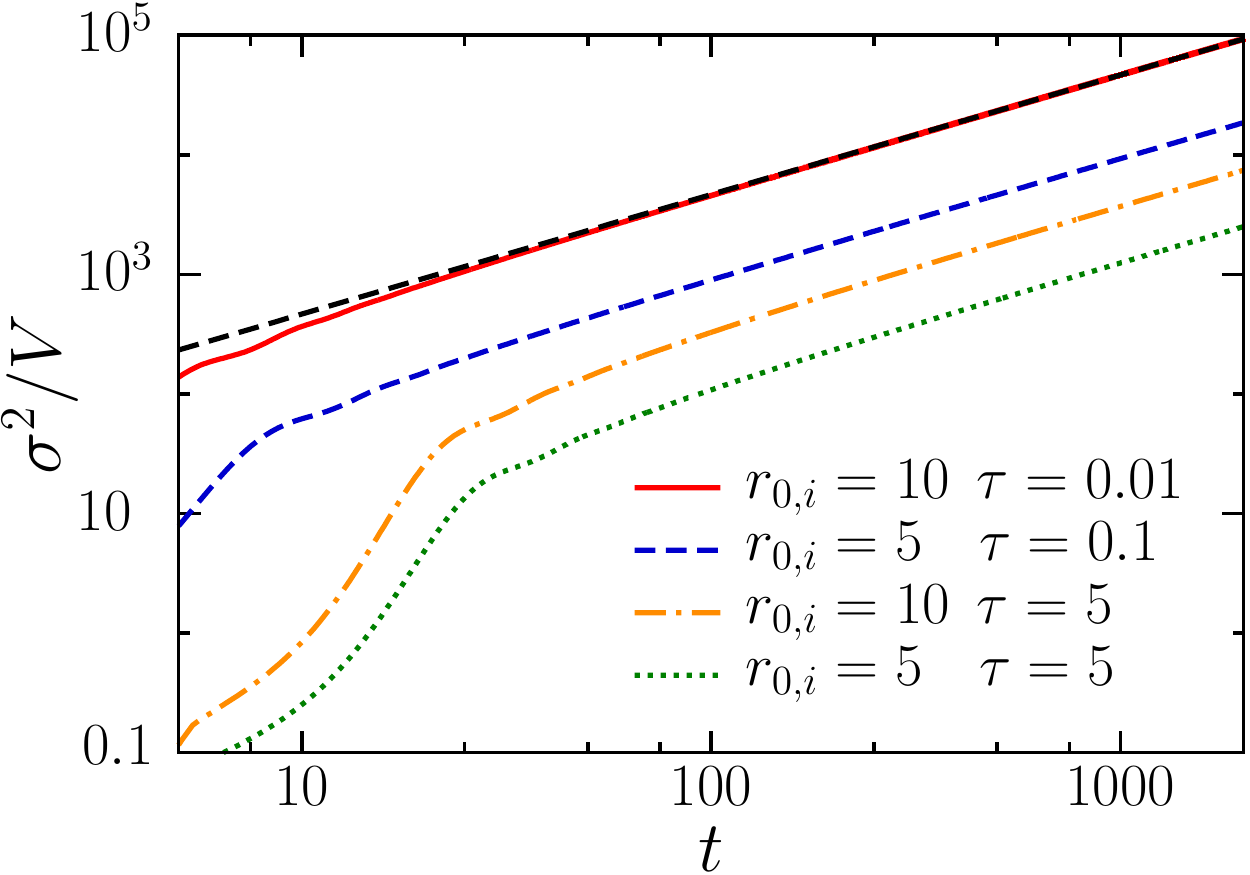}\label{fig:sigma3d<}}\\
\vspace{-0.1cm}
\subfigure[]{\includegraphics[width=.32\textwidth]{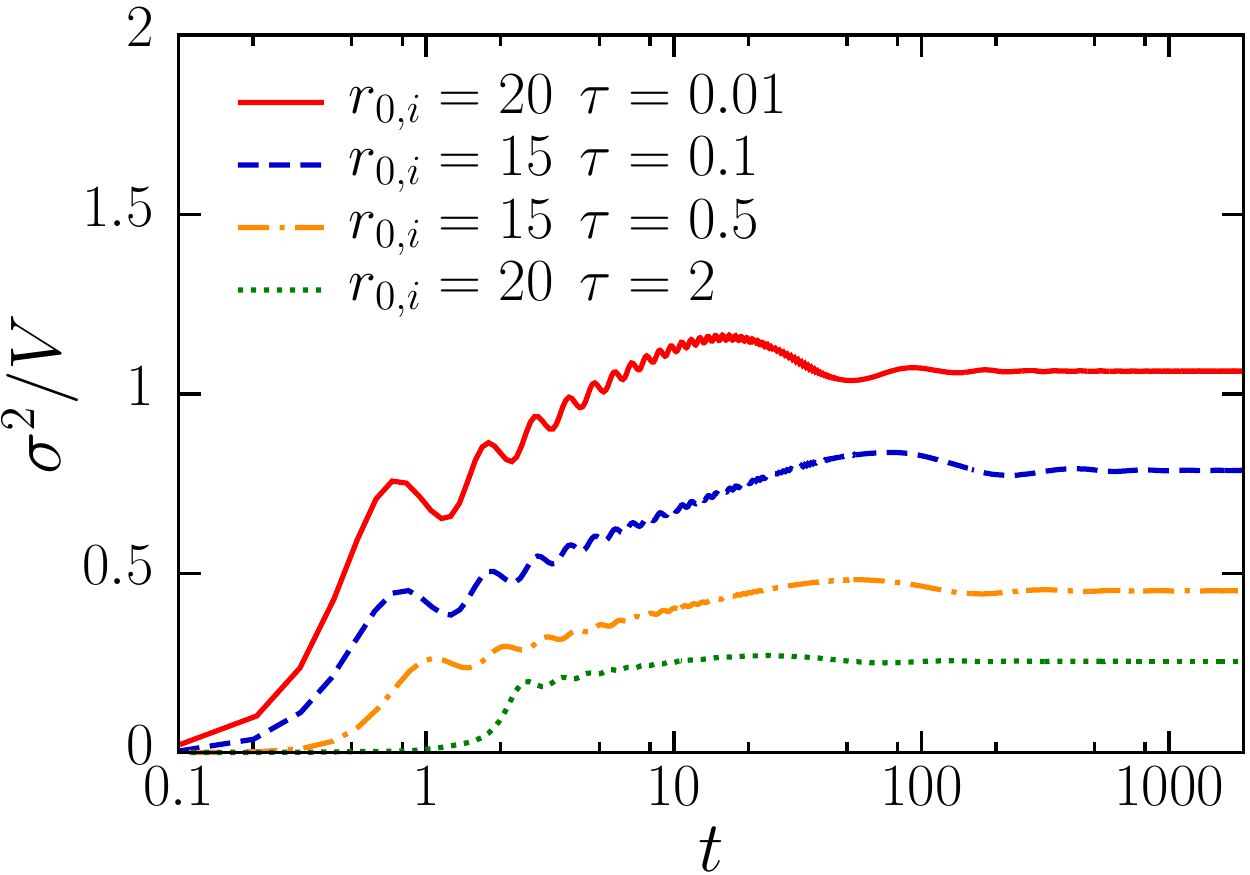}\label{fig:sigma4d>}}%
\subfigure[]{\hspace{0.27cm} \includegraphics[width=.32\textwidth]{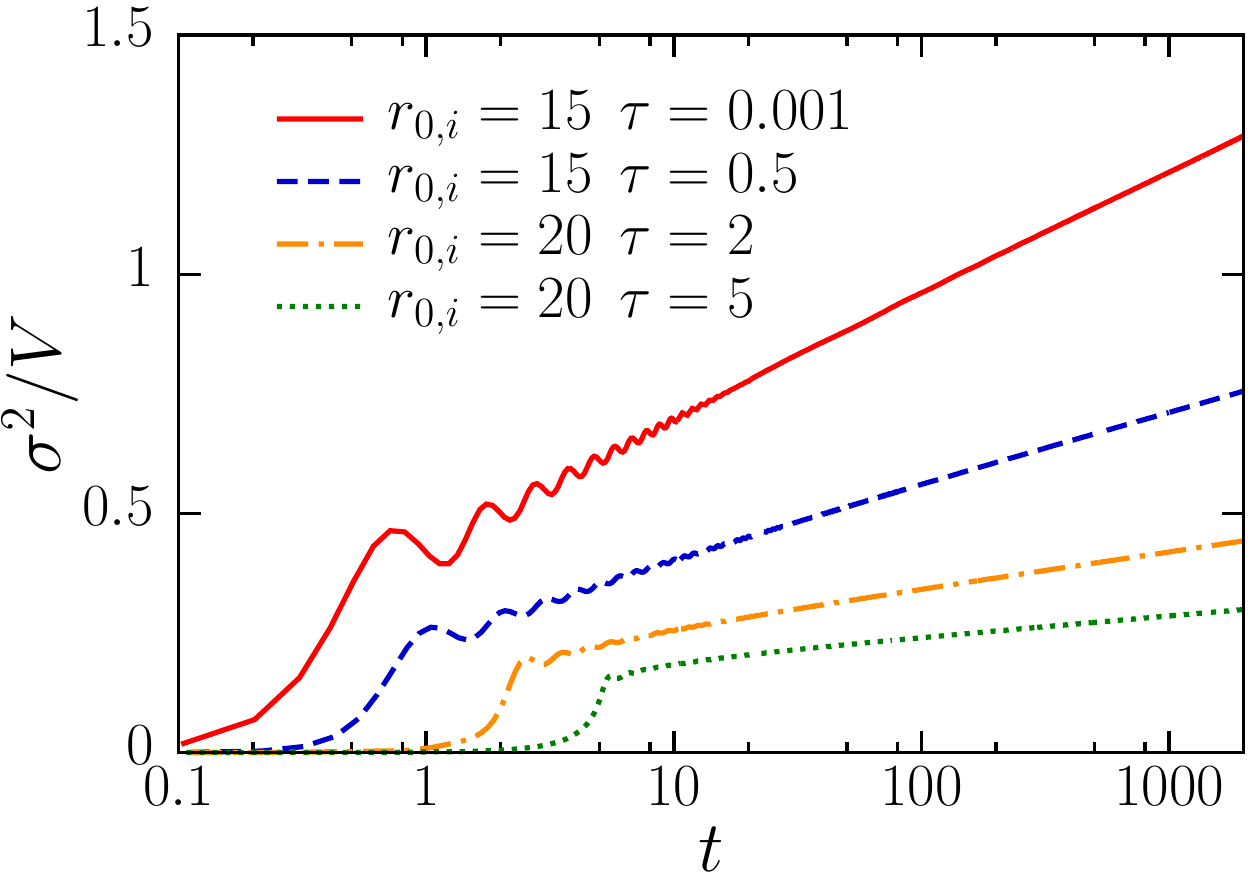}\label{fig:sigma4d=}}%
\subfigure[]{\hspace{0.27cm} \includegraphics[width=.32\textwidth]{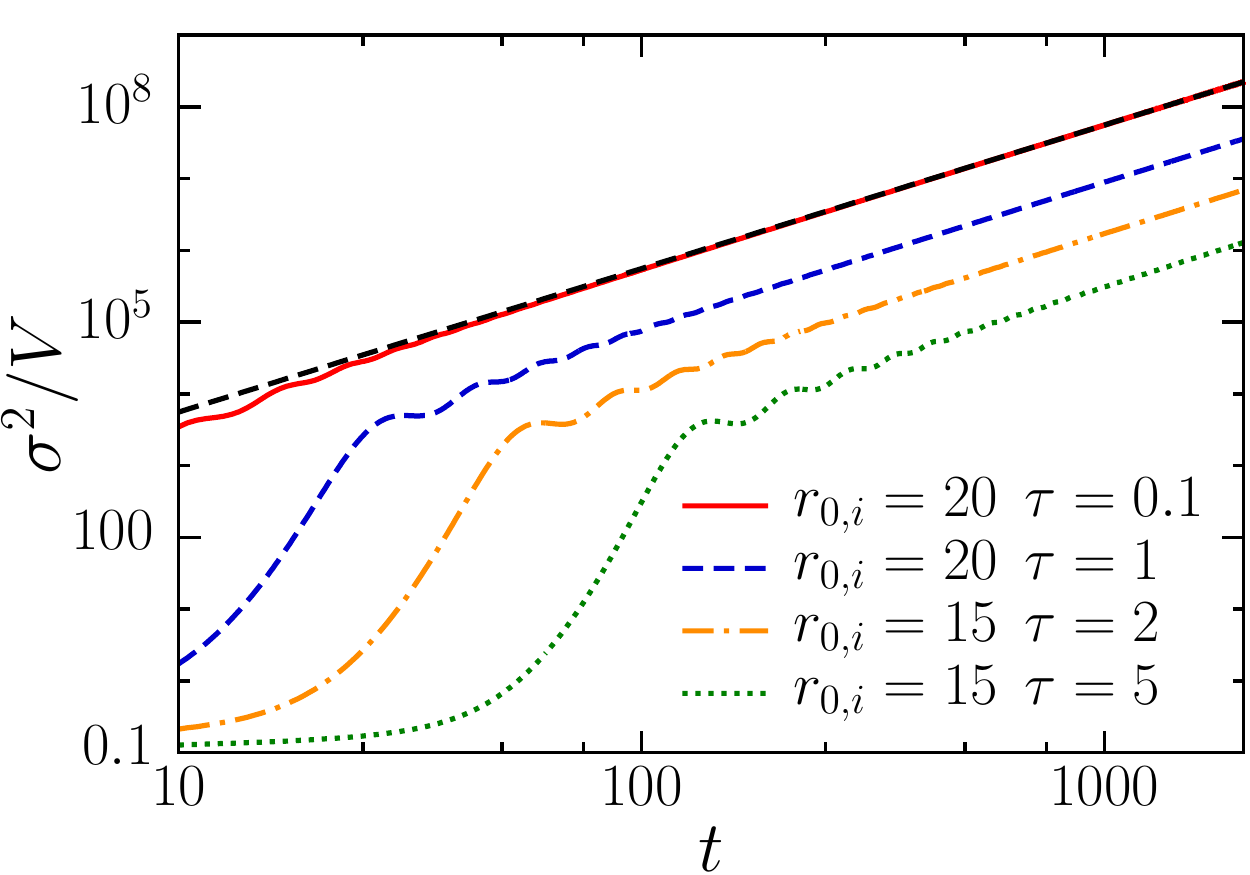}\label{fig:sigma4d<}}
\caption{(Color online) Variance per unit volume for ramps above~(a, d), at~(b, e), and below~(c, f) the dynamical critical point in $d=3$ (first row) and $d=4$ (second row). Different values of the initial bare mass $r_{0,i}$ and of the ramp duration $\tau$ are shown, while the interaction is $\lambda=15$. Black dashed lines in~\subref{fig:sigma3d<} and~\subref{fig:sigma4d<} are proportional respectively to $t$ and $t^2$.}%
\label{fig:sigma}
\end{figure*}

In order to complete our characterization of the crossover in the dynamical transition we study  the statistics of excitations produced by the ramp of the bare mass, generalizing the approach proposed in Ref.~\onlinecite{Smacchia2015}.
As in the case of a sudden quench, we will show that the growth in time of the fluctuations in the number of excitations bears strong signatures of the dynamical transition.

The protocol we will study is the following: 
after the end of the ramp, we let the system evolve for a certain waiting time after which we suddenly quench the bare mass back to its initial value $r_{0,i}$. 
The number of excitations generated in this process is a fluctuating quantity characterized by a certain probability distribution related to the operator
\beq
\hat{\mathcal{N}}=\int^{\Lambda} \!\!\! \frac{d^d k}{(2 \pi)^d} \, \ac_\bk \ad_\bk.
\eeq
An equivalent and more convenient description can be given in terms of the moment generating function
\beq
G(s, t)= \bra{\psi(t)} e^{-s\hat{\mathcal{N}}}  \ket{\psi(t)},
\eeq
where $\ket{\psi(t)}= U(t)\ket{0}$ is the evolved state at time $t$ and $\ket{0}$ indicates the initial ground state.
The explicit derivation of $G(s, t)$ is presented in Appendix~\ref{app:D}. In particular, we obtain 
\beq \label{eq:logG}
\ln G(s, t)  = - \frac{V}{2} \!  \int^{\Lambda} \!\!\! \frac{d^d k}{(2 \pi)^d}\, \ln \! \left[ 1+ \rho^{\phantom{\star}}_k(t) \! \left( 1-e^{-2s} \right) \right]\!,\!
\eeq
where
\beq \label{eq:rho}
\rho^{\phantom{\star}}_k(t) = \frac{1}{2} \left[ \omega_k(0) \lvert \fk (t) \rvert^2 + \frac{\lvert \dot{f}^{\phantom{\star}}_k(t) \rvert^2}{\omega_k(0)} -1 \right]
\eeq
and $V=L^d$, $L$ being the linear size of the system.

The dynamical critical properties of the system can be studied by analyzing the cumulants of the distribution of excitations, defined as
\beq
C_n(t)= (-1)^n \left. \frac{\partial^n}{\partial s^n} \ln G(s, t) \right|_{s=0}.
\eeq
In the following, we will focus on the first two cumulants, i.e., the average $\overline{\mathcal{N}}(t)$ and the variance $\sigma^2(t)$, in $d=3$ and $d=4$ and numerically study their time dependence, trying to distinguish qualitatively different behaviors for different values of the bare mass at the end of the ramp.
Their explicit expressions in terms of $\rho^{\phantom{\star}}_k(t)$ are
\begin{gather}
\frac{\overline{\mathcal{N}}(t)}{V} = \int^{\Lambda} \!\!\! \frac{d^d k}{(2 \pi)^d} \, \rho^{\phantom{\star}}_k(t), \\
\frac{\sigma^2(t)}{V}= \int^{\Lambda} \!\!\! \frac{d^d k}{(2 \pi)^d} \, 2 \rho^{\phantom{\star}}_k(t) \left[ 1+\rho^{\phantom{\star}}_k(t) \right].
\end{gather}

For large times, the average number of excitations relaxes to a finite value for every value of $r_{0,f}$, both in $d=3$ and $d=4$.
Remarkably, the variance per unit volume displays a non-trivial behavior at large times, depending on the final value of the bare mass $r_{0,f}$. For ramps ending above the dynamical critical point, i.e.,  $r_{0,f} > r_{0,f}^c(\tau)$, the variance saturates to a finite value, both in $d=3$ and in $d=4$ (Fig.~\ref{fig:sigma3d>} and~\ref{fig:sigma4d>}). For $r_{0,f} < r_{0,f}^c(\tau)$, the variance increases algebraically: for $d=3$ it scales as $\sigma^2 \sim t$ (Fig.~\ref{fig:sigma3d<}), while for $d=4$ it scales as $\sigma^2 \sim t^2$ (Fig.~\ref{fig:sigma4d<}). Finally, for ramps at the critical point, i.e.,  $r_{0,f} = r_{0,f}^c(\tau)$, the variance grows logarithmically in time, both in $d=3$ and in $d=4$ (Fig.~\ref{fig:sigma3d=} and~\ref{fig:sigma4d=}). 

We note that this behaviour is the same observed in the case of a sudden quench~\cite{Smacchia2015}, showing that the critical scaling of the variance appears to be unaffected by the change of the protocol.


\section{Linear ramp below the dynamical critical point}
\label{sec:Ramp-below}

\begin{figure*}[t]
\centering%
\subfigure[]{\includegraphics[width=.32\textwidth]{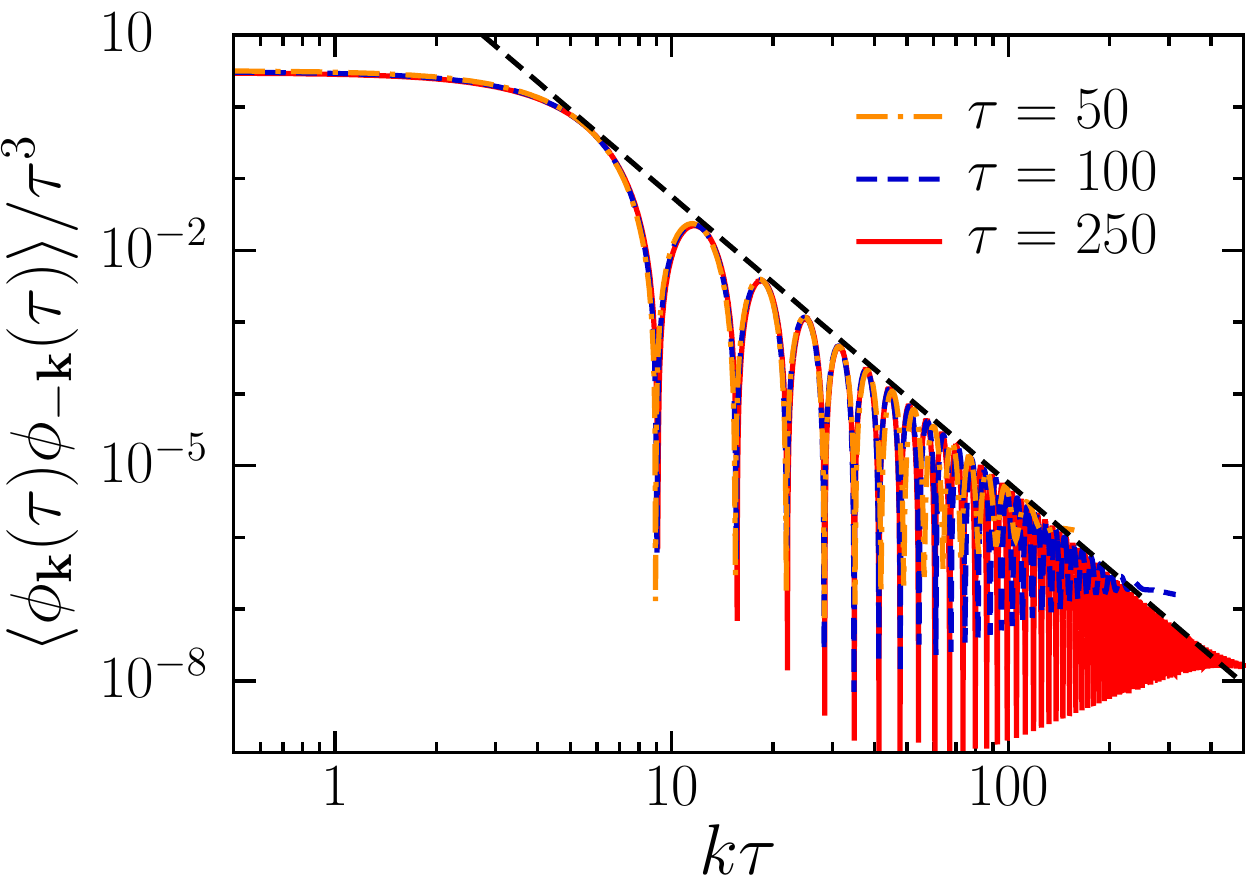}\label{fig:Gktau3d}}%
\subfigure[]{\hspace{0.27cm}\includegraphics[width=.32\textwidth]{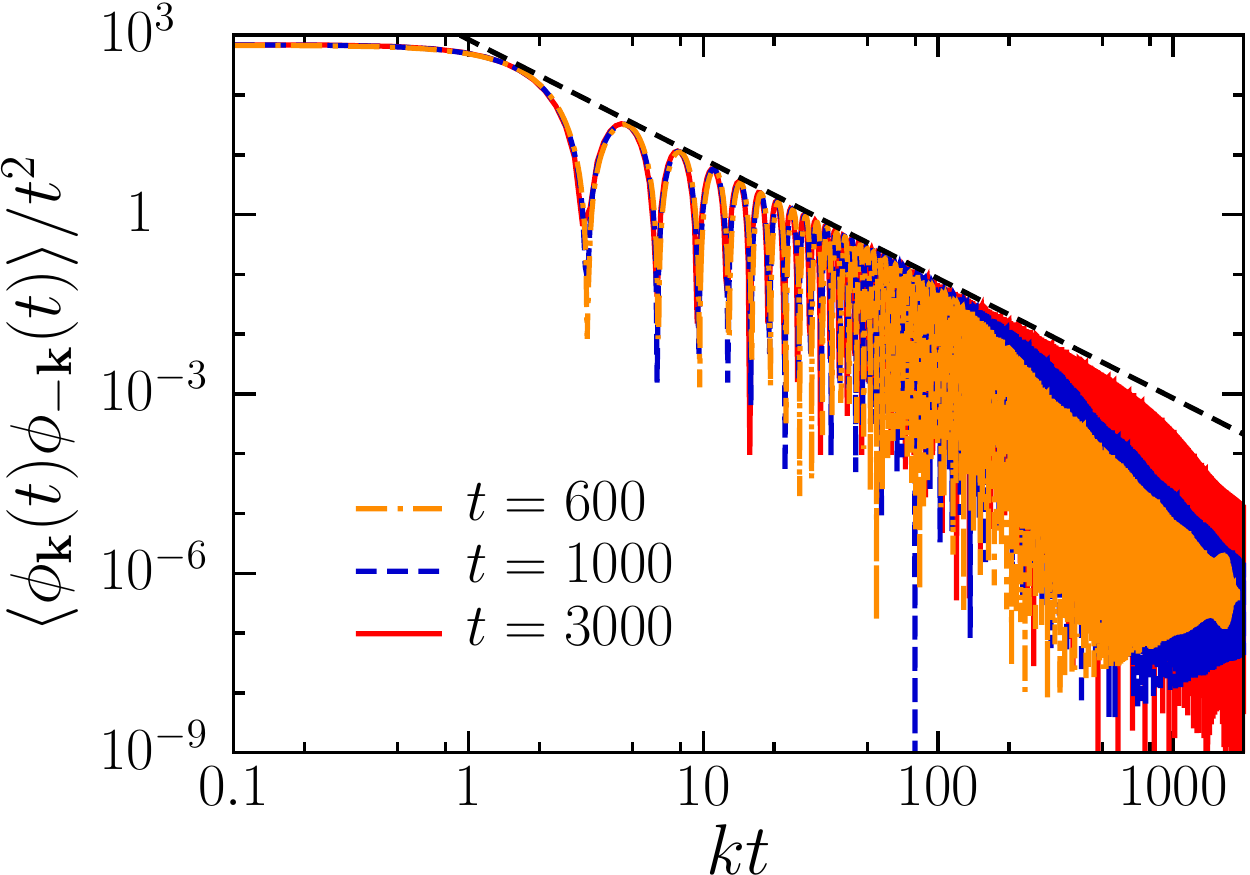}\label{fig:Gkt3d}}%
\subfigure[]{\hspace{0.27cm}\includegraphics[width=.32\textwidth]{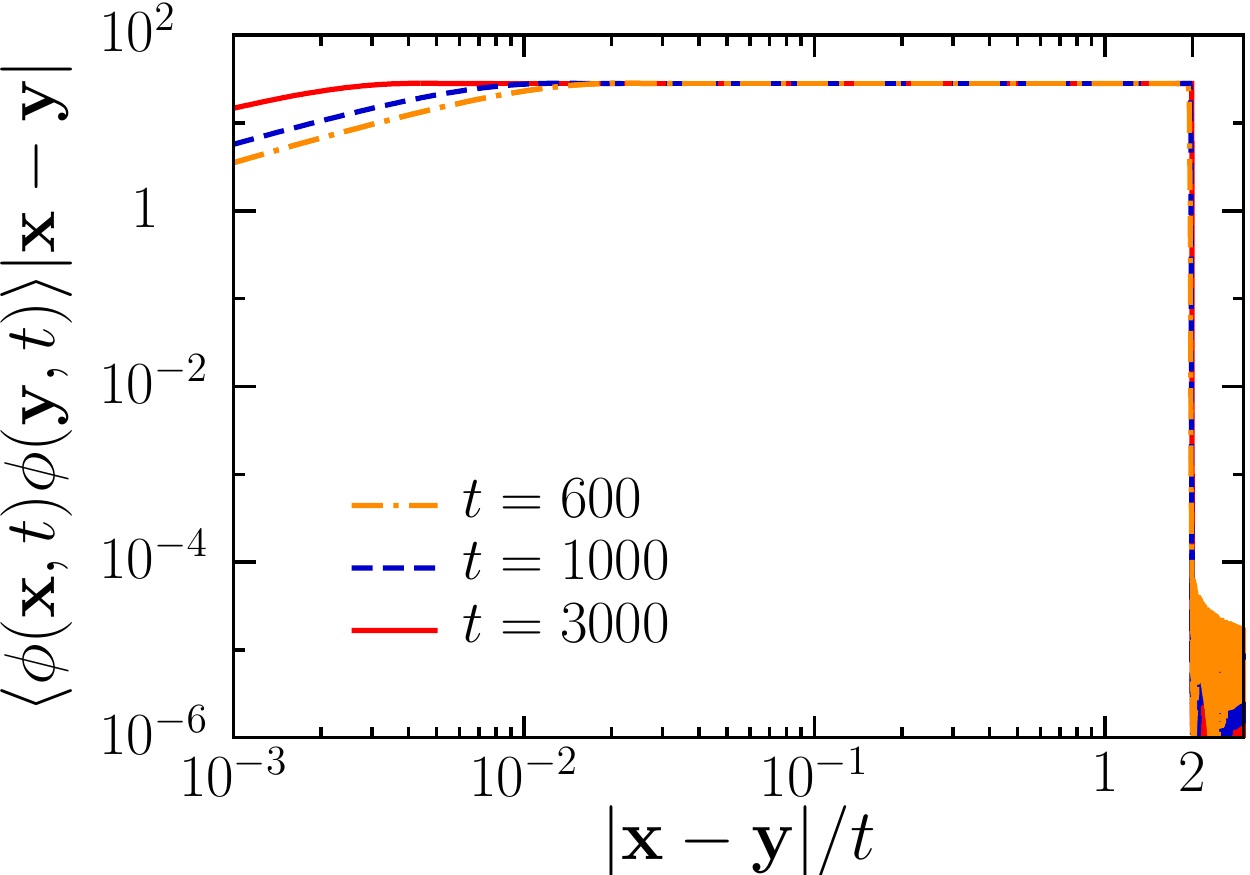}\label{fig:Gx3d}}\\
\vspace{-0.1cm}
\subfigure[]{\includegraphics[width=.32\textwidth]{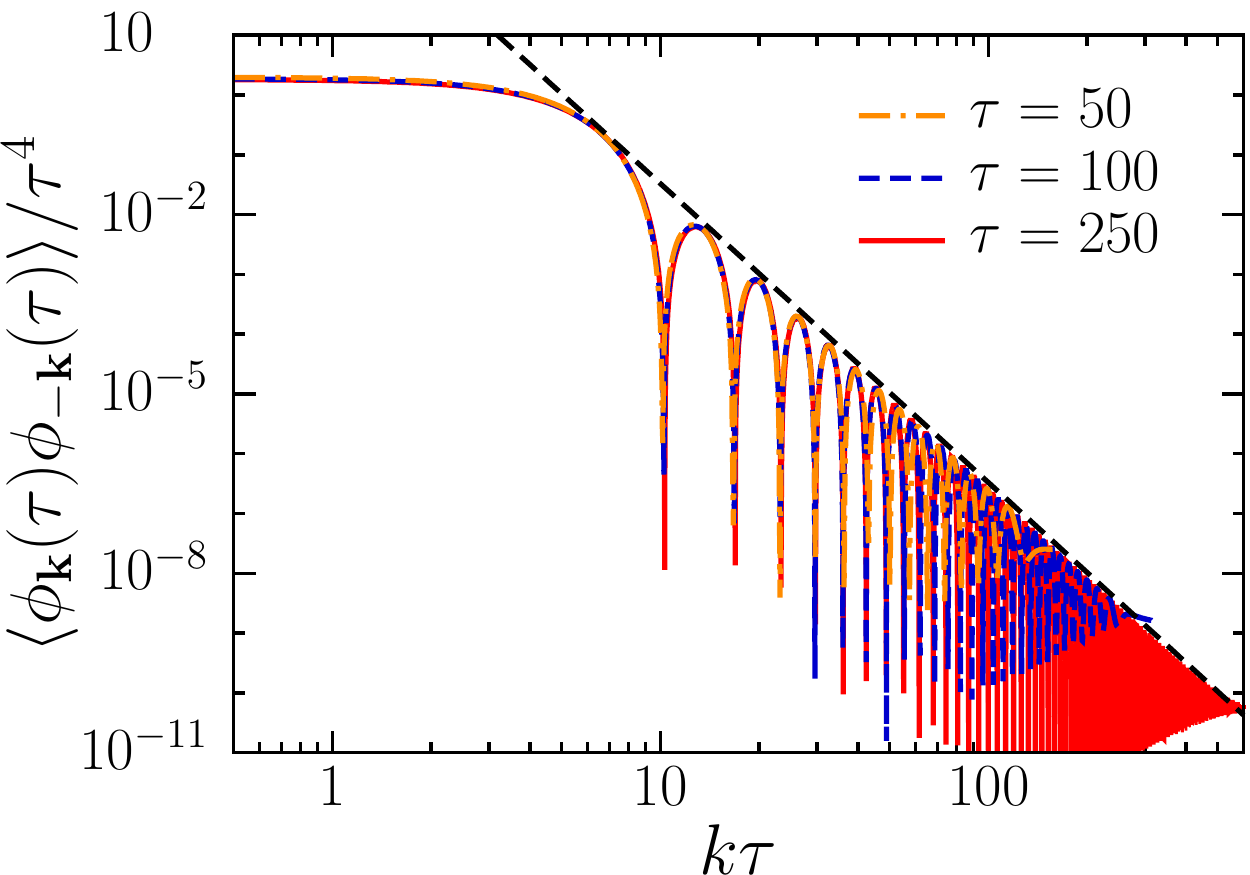}\label{fig:Gktau4d}}%
\subfigure[]{\hspace{0.27cm} \includegraphics[width=.32\textwidth]{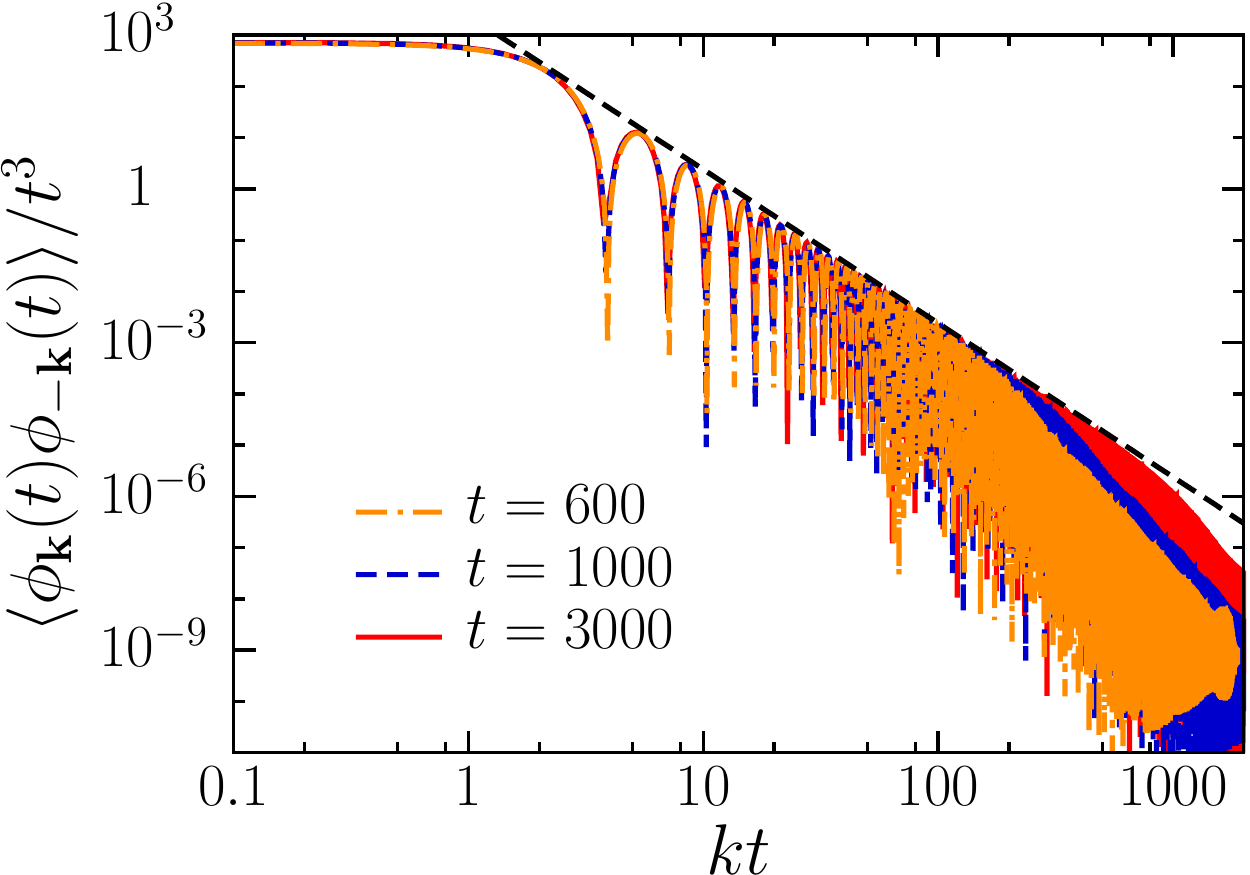}\label{fig:Gkt4d}}%
\subfigure[]{\hspace{0.27cm} \includegraphics[width=.32\textwidth]{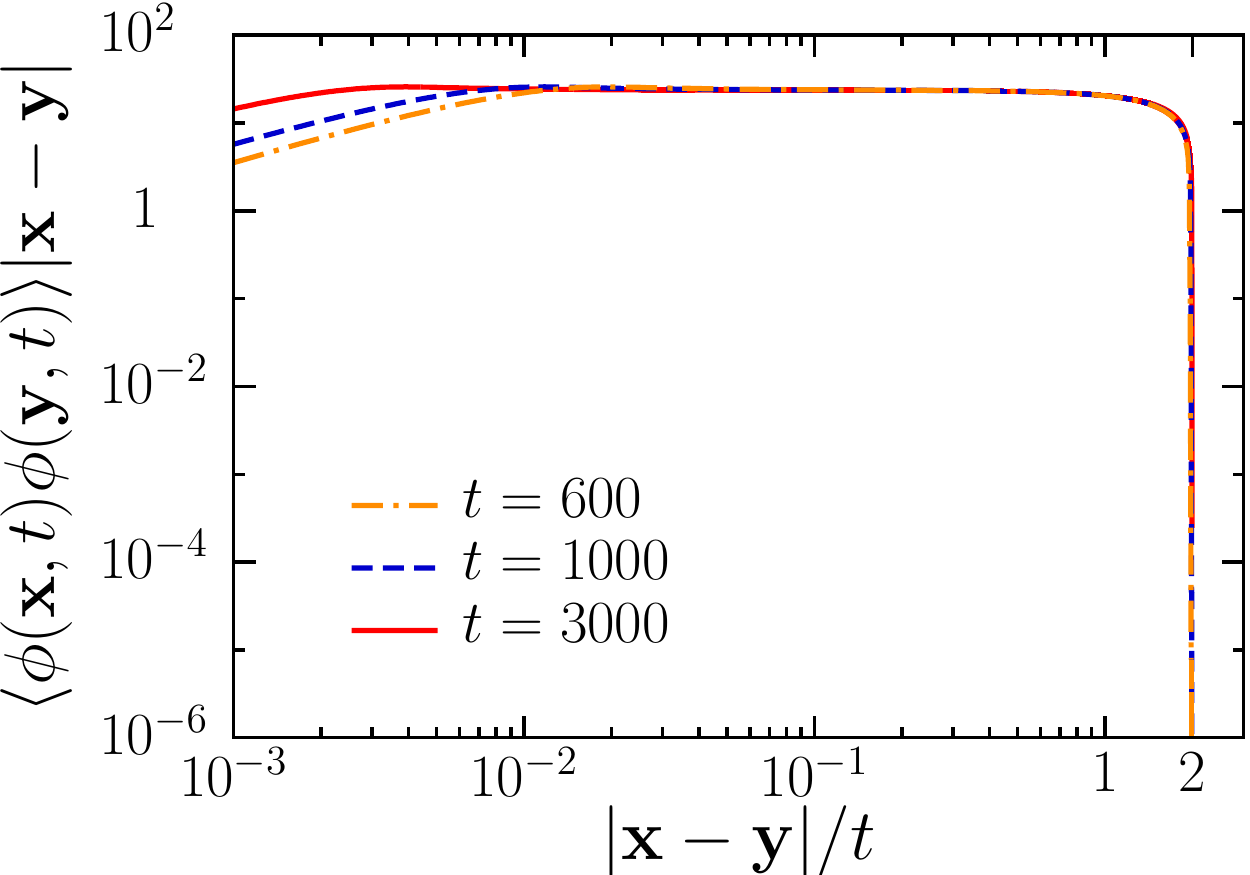}\label{fig:Gx4d}}
\caption{(Color online) Equal-time two-point correlation functions for ramps below the dynamical critical point in $d = 3$ (first row) and $d = 4$ (second row). In all the figures the initial bare mass is $r_{0,i}=15$, the final bare mass is~$r_{0,f}=-15$ and the interaction is $\lambda=15$. %
In~\subref{fig:Gktau3d} and~\subref{fig:Gktau4d} different values of the ramp duration $\tau$ are shown and the black dashed lines are proportional respectively to $(k \tau)^{-4}$ and $(k \tau)^{-5}$. %
In~\subref{fig:Gkt3d} and~\subref{fig:Gkt4d} different values of the evolution time $t$ are shown, for a ramp of duration~$\tau=20$, and the black dashed lines are proportional respectively to $(k t)^{-2}$ and $(k t)^{-3}$. %
In~\subref{fig:Gx3d} and~\subref{fig:Gx4d} different values of the evolution time $t$ are shown, for a ramp of duration~$\tau=20$.}%
\label{fig:below}
\end{figure*}

An interesting signature of the crossover between quench and slow ramp is observed by focusing on 
ramps below the dynamical critical point.
It has been shown~\cite{Chandran2013, sciolla_2013} that performing a sudden quench below the dynamical critical point induces the emergence of a scaling form in the correlation functions associated with coarsening dynamics with an exponent characterizing these functions differing from the one expected in usual classical coarsening.
The reason for this discrepancy between quantum and classical systems is till now unclear.
We investigate how this behavior is affected by a linear ramp in the bare mass.
To this end, we consider the equal-time two-point correlation function $\langle \phi_\bk(t) \phi_{-\bk}(t)\rangle=\lvert \fk (t) \rvert^2$ and its Fourier transform $\langle \phi(\mathbf{x}, t) \phi(\mathbf{y}, t)\rangle$  in $d=3$ and $d=4$.

As a consequence of the ramp protocol, the dependence of~$\langle \phi_\bk(t) \phi_{-\bk}(t)\rangle$ on momentum~$k$ displays two different regimes.
Right at the end of the ramp (Fig.~\ref{fig:Gktau3d} and~\ref{fig:Gktau4d}), we note that it exhibits the following scaling form:
\beq
\langle \phi_\bk(\tau) \phi_{-\bk}(\tau)\rangle = \tau^d \mathcal{F}_d (k \tau),
\eeq
where~$\mathcal{F}_d (k \tau)$ is an oscillating function decaying as~$\sim (k \tau)^{-(d+1)}$ for~$k \tau \gtrsim 1$.

In the subsequent evolution for~$t>\tau$, shown in Fig.~\ref{fig:Gkt3d} and~\ref{fig:Gkt4d}, the correlation function acquires a different dependence on momentum for~$1/t \lesssim k \lesssim 1/\tau$ and, for long times after the end of the ramp, the scaling form found in the case of a sudden quench is recovered, namely,
\beq
\langle \phi_\bk(t) \phi_{-\bk}(t)\rangle = t^{d-1} \mathcal{G}_d (k t),
\eeq 
where~$\mathcal{G}_d (k t)$ is an oscillating function decaying as~$\sim (k t)^{-(d-1)}$ for~$1 \lesssim kt \lesssim t/\tau$.
For~$k \gtrsim 1/\tau $, instead, the correlation function still decays as~$\sim k^{-(d+1)}$.
Notice that in the limit~$\tau \to 0$ the latter regime, which is due to the finite duration of the ramp, is suppressed.

The corresponding Fourier transform in real space,~$\langle \phi(\mathbf{x}, t) \phi(\mathbf{y}, t)\rangle$, shown in Fig.~\ref{fig:Gx3d} and~\ref{fig:Gx4d}, exhibits a light-cone structure~\cite{calabrese_07, cheneau_2012, sciolla_2013}, vanishing for~$\lvert \mathbf{x}-\mathbf{y} \rvert > 2t$ as a consequence of the finite speed of propagation of excitations, and it decays as $\lvert \mathbf{x}-\mathbf{y} \rvert^{-1}$ for $\tau \lesssim \lvert \mathbf{x}-\mathbf{y} \rvert < 2t$, both in $d=3$ and in $d=4$.
While in the limit~$\tau \to 0$ the result of a sudden quench is fully recovered, for~$\tau \to \infty$ we do not find the corresponding equilibrium correlation function, since the $O(N)$ symmetry can not be globally broken by the dynamics.
Moreover, adiabaticity is not expected to hold, since the system crosses the dynamical critical point and enters a gapless phase.

%


\section{Conclusions}
\label{sec:Conlusions}

In this work we investigated the crossover of the dynamical phase transitions of the $O(N)$ vector model in the $N \to \infty$ limit as a function of the duration
of a  linear ramp in the bare mass. In particular, we have shown that, when  the duration of the ramp is finite, the critical properties associated to dynamical transitions 
are the same as the equilibrium transition at finite temperature, while as $\tau \rightarrow +\infty$ they are close to the ones of the equilibrium system at zero temperature, i.e. the quantum phase transition. Studying in detail the location of the dynamical critical point $r_{0,f}^c(\tau)$, we investigated how its value interpolates between the limiting cases of the sudden quench ($\tau \to 0$) and the adiabatic switching ($\tau \to \infty$) of the bare mass. We found that the approach to these two limits is algebraic in $\tau$ and we derived analytically the values of such exponents.
 
As for a quench, the nonequilibrium nature of the dynamical transition, however, leaves strong signatures on the statistics of the excitations, whose variance
grows as a power law below the critical point and exhibits a logarithmic behaviour at the critical point.
An intriguing crossover is finally observed analyzing the equal-time two-point correlation function for ramps below the dynamical critical point. There we found the emergence of two different scaling behaviors, one related to the finite duration of the ramp (unrelated to quantum critical scaling) and the other to the subsequent time evolution and coarsening dynamics.


%
%


\appendix

\section{Non-interacting theory and stationary values}
\label{app:A}

In this section we explicitly solve Eqs.~(\ref{eq:f-r}) in the case of a free theory, i.e., $\lambda=0$, and give additional details on the ansatz of Eq.~(\ref{eq:r*_ramp}).

Obviously, when $\lambda=0$ the effective mass is equal to the bare one, therefore we have to solve the equation 
\beq 
\ddot{f}^0_k (t) +\left( k^2+r_0(t) \right) f^0_k (t) = 0
\eeq
with initial conditions $f^0_k (0)=1/(4(k^2+r_{0,i}))^{1/4}$ and $\dot{f}^0_k(0)=-i((k^2+r_{0,i})/4)^{1/4}$. 

For $0<t<\tau$, the solution is given by
\bw
\begin{align} \label{eq:f0<}
f^0_k (t) &= \frac{\pi}{\sqrt{2}(k^2+r_{0,i})^{1/4}} \left[ \Ai\left( \gamma t -\frac{k^2+r_{0,i}}{\gamma^2} \right) \Bi'\left(-\frac{k^2+r_{0,i}}{\gamma^2} \right) -\Ai'\left(-\frac{k^2+r_{0,i}}{\gamma^2} \right) \Bi\left( \gamma t -\frac{k^2+r_{0,i}}{\gamma^2} \right) \right] \nonumber \\
& + \frac{i \pi (k^2+r_{0,i})^{1/4}}{\sqrt{2} \gamma} \left[ \Ai\left( \gamma t -\frac{k^2+r_{0,i}}{\gamma^2} \right) \Bi\left(-\frac{k^2+r_{0,i}}{\gamma^2} \right) -\Ai\left(-\frac{k^2+r_{0,i}}{\gamma^2} \right) \Bi\left( \gamma t -\frac{k^2+r_{0,i}}{\gamma^2} \right) \right],
\end{align}
where $\gamma=((r_{0,i}-r_{0,f})/\tau)^{1/3}$, and $\Ai(x)$ and $\Bi(x)$ denote the Airy functions, while for $t>\tau$
\beq \label{eq:f0>}
f^0_k (t) = f^0_k (\tau) \cos \left(\sqrt{k^2+r_{0,f}} (t-\tau)\right) + \frac{\dot{f}^0_k(\tau)}{\sqrt{k^2+r_{0,f}}} \sin \left(\sqrt{k^2+r_{0,f}} (t-\tau) \right),
\eeq
where $f^0_k (\tau)$ and $\dot{f}^0_k(\tau)$ have to be read from Eq.~(\ref{eq:f0<}).

Using Eqs.~(\ref{eq:f0<}) and~(\ref{eq:f0>}), we can explicitly compute the two-body equal time Green's function $\langle \phi_\bk(t) \phi_{-\bk}(t)\rangle=\lvert f^0_k(t) \rvert^2$, that, for $t>\tau$, is
\begin{align} \label{eq:f^2}
\langle \phi_\bk(t) \phi_{-\bk}(t)\rangle = \frac{1}{2} &\left[\lvert f^0_k (\tau) \rvert^2 + \frac{\lvert \dot{f}^0_k (\tau) \rvert^2}{k^2 + r_{0,f}} + \left( \lvert f^0_k (\tau) \rvert^2 - \frac{\lvert \dot{f}^0_k (\tau) \rvert^2}{k^2 + r_{0,f}} \right) \cos \left( \sqrt{k^2+r_{0,f}} (t-\tau) \right) \right. \nonumber \\
&\left. + \frac{2 \Rea \left( f^0_k (\tau) \dot{f}^{0 \star}_k (\tau) \right)}{\sqrt{k^2+r_{0,f}}} \sin \left( \sqrt{k^2+r_{0,f}} (t-\tau) \right) \right]. 
\end{align}
\ew

When $\lambda \neq 0$, we have to resort to numerical integration of Eqs.~(\ref{eq:f-r}), which shows that for long times after the end of the ramp the effective mass $r(t)$ relaxes to a stationary value. 
In order to predict the stationary value $r^*$, we use the following ansatz: after the end of the ramp, we assume the stationary part of the two-body equal time Green's function to be equal to the non-interacting one, but with the bare masses and the ramp duration replaced by the renormalized ones. Namely, we take Eq.~(\ref{eq:f^2}), disregard all the oscillatory terms and replace $r_{0,i}\rightarrow r_i$, $r_{0,f}\rightarrow r^*$ and $\tau \rightarrow \tilde{\tau}$.
Thus, we obtain the following self-consistent equation for $r^*$
\beq \label{eq:r*_ramp_app}
r^* \! =r_{0,f}+ \frac{\lambda}{12} \int^{\Lambda} \!\!\! \frac{d^d k}{(2 \pi)^d} \! \left[ \lvert f^0_k (r^*\!, \tilde{\tau}) \rvert^2 + \frac{\lvert \dot{f}^0_k (r^*\!, \tilde{\tau}) \rvert^2}{k^2 + r^*} \right] \!,\!\!
\eeq
where
\bw
\begin{align} \label{mode1}
f^0_k (r^*\!, \tilde{\tau}) =& \frac{\pi}{\sqrt{2}(k^2+r_i)^{1/4}} \left[ \Ai\left(-\frac{k^2+r^*}{\tilde{\gamma}^2} \right) \Bi'\left(-\frac{k^2+r_i}{\tilde{\gamma}^2} \right) -\Ai'\left(-\frac{k^2+r_i}{\tilde{\gamma}^2} \right) \Bi\left(-\frac{k^2+r^*}{\tilde{\gamma}^2} \right) \right] \nonumber \\
& + \frac{i \pi (k^2+r_i)^{1/4}}{\sqrt{2} \tilde{\gamma}} \left[ \Ai\left(-\frac{k^2+r^*}{\tilde{\gamma}^2} \right) \Bi\left(-\frac{k^2+r_i}{\tilde{\gamma}^2} \right) -\Ai\left(-\frac{k^2+r_i}{\tilde{\gamma}^2} \right) \Bi\left(-\frac{k^2+r^*}{\tilde{\gamma}^2} \right) \right], \\
\dot{f}^0_k (r^*\!, \tilde{\tau}) =& \frac{\pi \tilde{\gamma}}{\sqrt{2}(k^2+r_i)^{1/4}} \left[ \Ai'\left(-\frac{k^2+r^*}{\tilde{\gamma}^2} \right) \Bi'\left(-\frac{k^2+r_i}{\tilde{\gamma}^2} \right) -\Ai'\left(-\frac{k^2+r_i}{\tilde{\gamma}^2} \right) \Bi'\left(-\frac{k^2+r^*}{\tilde{\gamma}^2} \right) \right] \nonumber \\
& + \frac{i \pi (k^2+r_i)^{1/4}}{\sqrt{2}} \left[ \Ai'\left(-\frac{k^2+r^*}{\tilde{\gamma}^2} \right) \Bi\left(-\frac{k^2+r_i}{\tilde{\gamma}^2} \right) -\Ai\left(-\frac{k^2+r_i}{\tilde{\gamma}^2} \right) \Bi'\left(-\frac{k^2+r^*}{\tilde{\gamma}^2} \right) \right],\label{mode2}
\end{align}
\ew
with $\tilde{\gamma}=((r_i-r^*)/\tilde{\tau})^{1/3}$.


\section{Dynamical critical properties}
\label{app:B}

In this section we provide the detailed computation of the critical dimensions and critical exponent $\nu^*$.

For studying the lower critical dimension, it is useful to remind the expansions of the Airy functions both for small and large arguments~\cite{abramowitz}.
For small $x$ we have
\begin{subequations}\label{eq:airy_small}
\begin{align}
&\Ai(-x)= \frac{1}{3^{2/3} \Gamma(2/3)} + \frac{x}{3^{1/3} \Gamma(1/3)} + O(x^3), \\
&\Ai'(-x)= -\frac{1}{3^{1/3} \Gamma(1/3)} + \frac{x^2}{2 \! \cdot \! 3^{2/3} \Gamma(2/3)} + O(x^3), \!\!\! \\
&\Bi(-x)= \frac{1}{3^{1/6} \Gamma(2/3)} - \frac{3^{1/6} x}{\Gamma(1/3)} + O(x^3), \\
&\Bi'(-x)= \frac{3^{1/6}}{\Gamma(1/3)} + \frac{x^2}{2 \! \cdot \! 3^{1/6} \Gamma(2/3)} + O(x^3),
\end{align}
\end{subequations}
while for large and positive $x$ we have
\begin{subequations} \label{eq:airy_large}
\begin{align}
&\Ai(-x)= \frac{1}{\sqrt{\pi} x^{1/4}} \sin \left( \frac{\pi}{4} + \frac{2}{3} x^{3/2} \right) + O(x^{-7/4}), \! \\
&\Ai'(-x)= - \frac{x^{1/4}}{\sqrt{\pi}} \cos \left( \frac{\pi}{4} + \frac{2}{3} x^{3/2} \right) + O(x^{-5/4}), \\
&\Bi(-x)= \frac{1}{\sqrt{\pi} x^{1/4}} \cos \left( \frac{\pi}{4} + \frac{2}{3} x^{3/2} \right) + O(x^{-7/4}), \! \\
&\Bi'(-x)= \frac{ x^{1/4}}{\sqrt{\pi}} \sin \left( \frac{\pi}{4} + \frac{2}{3} x^{3/2} \right) + O(x^{-5/4}).
\end{align}
\end{subequations}
We can now analyze the behavior for low momenta of the integrand of Eq.~(\ref{eq:r*_ramp_app}) with $r^*=0$.
For every finite $\tau$, the most relevant modes are those with $k \ll (r_i/\tilde{\tau})^{1/3}$ and $k \ll \sqrt{r_i}$.
In this region and for $r^*=0$, we can replace all the Airy functions with argument $-(k^2+r^*)/\tilde{\gamma}^2$ with their zero value (see Eqs. (\ref{eq:airy_small})), while the leading order of all the other terms are obtained setting $k=0$.
Thus, we conclude that both $f^0_k (0, \tilde{\tau})$ and $\dot{f}^0_k (0, \tilde{\tau})$  are constant in $k$.
As a consequence, the lower critical dimension for every finite $\tau$ is $d=2$.
Instead, to understand what happens in the limit $\tau \to \infty$, we have to take into account the region $(r_i/\tilde{\tau})^{1/3} \ll k \ll \sqrt{r_i}$. 
Here, and for $r^*=0$, we have to substitute the Airy functions with argument $-(k^2+r^*)/\tilde{\gamma}^2$ with their asymptotic expansions of Eqs.~(\ref{eq:airy_large}), and set $k=0$ in all the other terms. 
Thus, we see that $f^0_k (0, \tilde{\tau}) \sim 1/\sqrt{k}$ and $\dot{f}^0_k (0, \tilde{\tau}) \sim \sqrt{k}$.
So, when $\tau$ is strictly infinite the lower critical dimension is $d = 1$.

In order to determine the critical exponent $\nu^*$, we analyze the behavior of the asymptotic mass $r^*$ for small distances of $r_{0,f}$ from the dynamical critical point.
Denoting $\delta r _{0, f}(\tau)= r_{0,f} - r_{0,f}^c(\tau)$, defining the dimensionless variable ${\bf{y}}=\bk / \sqrt{r^*}$, and combining Eqs.~(\ref{eq:r*_ramp}) and~(\ref{eq:rc_ramp}), we can write 
\bw
\beq \label{eq:r*_y}
r^*\! = \delta r _{0, f}(\tau) + \frac{\lambda}{12}(r^*)^{\frac{d-2}{2}} \int^{\Lambda/\sqrt{r^*}} \!\! \frac{d^d y}{(2 \pi)^d} \, \frac{y^2 g(y \sqrt{r^*}, r^*) - (y^2+1) g(y \sqrt{r^*}, 0)}{y^2(y^2+1)},
\eeq
\ew
with
\beq
g(k, r^*)= \lvert f^0_k (r^*\!, \tilde{\tau}) \rvert^2 (k^2 + r^*) + \lvert \dot{f}^0_k (r^*\!, \tilde{\tau}) \rvert^2.
\eeq
The asymptotic behavior of the integral in Eq.~(\ref{eq:r*_y}) for small $r^*$ is determined by the behavior of the integrand in the region $1 \ll y \ll \sqrt{r_i/r^*}$, where it scales as $g(0, 0)/y^4$.
Thus, for $2<d<4$ the dominant contribution to the integral in powers of $r^*$ is obtained by replacing the upper limit of integration with infinity and the integrand with its leading order in $r^*$, namely
\bw
\beq
r^*\! \simeq \delta r _{0, f}(\tau) - \frac{\lambda}{12} \frac{\Omega(d)}{(2\pi)^d}(r^*)^{\frac{d-2}{2}} \int_0^\infty \!\! d y \, y^{d-1} \frac{g(0, 0)}{y^2(y^2+1)}=  \delta r _{0, f}(\tau) + \frac{\lambda}{12} \frac{\Omega(d)}{(2\pi)^d}(r^*)^{\frac{d-2}{2}} \frac{\pi \, g(0, 0)}{2 \sin\left( d \pi/2 \right)},
\eeq
\ew
where $\Omega(d)$ is the solid angle.
So, we conclude that at the leading order $r^* \sim (\delta r_{0, f}(\tau))^\frac{2}{d-2}$. 
For $d=4$ we have logarithmic corrections to this scaling, while for $d>4$ the divergence of the integral can be deduced by considering the scaling of the integrand mentioned above. 
We have that the integral diverges as $(r^*)^{-\frac{d-4}{2}}$, giving a linear relation $r^* \sim \delta r_{0, f}(\tau)$ at the leading order.
Therefore, we can recover the values of Eq.~(\ref{eq:nu*}) for the critical exponent~$\nu^*$.


\section{Asymptotic expansions for large $\tau$}
\label{app:C}

In this section we provide additional details on the derivation of Eqs.~(\ref{eq:rc_large-3d}) and~(\ref{eq:rc_large-4d}).

Let us start by considering the case of $d=3$.
Computing the integrals~(\ref{eq:I}), we obtain
\bw
\begin{subequations}
\begin{align}
\mathcal{I}_1(3)= \Lambda^2 &\left[ \frac{2 \pi}{3^{1/3} \Gamma^2 (-1/3)}\, {_2}F_3 \! \left( \frac{1}{6}, \frac{1}{2}; \frac{1}{3}, \frac{2}{3}, \frac{3}{2}; -\frac{4}{9} y^2 \right) y^{1/3} - \frac{1}{5 \sqrt{3}}\, {_2}F_3 \!\left( \frac{1}{2}, \frac{5}{6}; \frac{2}{3}, \frac{4}{3}, \frac{11}{6}; -\frac{4}{9} y^2 \right) y  \right. \nonumber \\
& \left. + \frac{\Gamma(5/6)}{2^{1/3} \! \cdot \! 3^{1/6} \! \cdot \! 7 \sqrt{\pi}} \, {_2}F_3 \!\left( \frac{5}{6}, \frac{7}{6}; \frac{4}{3}, \frac{5}{3}, \frac{13}{6}; -\frac{4}{9} y^2 \right) y^{5/3} \right], \\
\mathcal{I}_2(3)= \Lambda^2 &\left[ -\frac{3^{1/3} \Gamma(-1/3) \Gamma(5/3)}{4 \pi}\, {_2}F_3 \! \left(\! -\frac{1}{6}, \frac{1}{6}; -\frac{1}{3}, \frac{1}{3}, \frac{7}{6}; -\frac{4}{9} y^2 \right) y^{-1/3} + \frac{1}{10 \sqrt{3}}\, {_2}F_3 \!\left( \frac{1}{2}, \frac{5}{6}; \frac{1}{3}, \frac{5}{3}, \frac{11}{6}; -\frac{4}{9} y^2 \right) y  \right. \nonumber \\
& \left. + \frac{\Gamma(1/6)}{2^{2/3} \! \cdot \! 3^{5/6} \! \cdot \! 36 \sqrt{\pi}} \, {_2}F_3 \!\left( \frac{7}{6}, \frac{3}{2}; \frac{5}{3}, \frac{7}{3}, \frac{5}{2}; -\frac{4}{9} y^2 \right) y^{7/3} \right],
\end{align}
\end{subequations}
where ${_2}F_3(a,b;c,d,e;x)$ denotes the hypergeometric function and $y=\Lambda^3 \tilde{\tau}/r_i$.
Taking the asymptotic expansions of the hypergeometric functions for large $y$, namely
\begin{subequations}
\begin{align}
&{_2}F_3 \! \left( \frac{1}{6}, \frac{1}{2}; \frac{1}{3}, \frac{2}{3}, \frac{3}{2}; -\frac{4}{9} y^2 \right)= \frac{3^{5/6} \sqrt{\pi}}{2^{1/3} \Gamma(1/6)} \, y^{-1/3} - \frac{\sqrt{3 \pi}\, \Gamma(-1/3)}{24 \,\Gamma(1/6)} \, y^{-1} + O(y^{-5/3}),\\
&{_2}F_3 \!\left( \frac{1}{2}, \frac{5}{6}; \frac{2}{3}, \frac{4}{3}, \frac{11}{6}; -\frac{4}{9} y^2 \right)= \frac{5}{4 \sqrt{3}} \, y^{-1} - \frac{5 \, \Gamma(-1/3)}{2^{8/3} \! \cdot \! 3^{11/6}} \, y^{-5/3} + O(y^{-7/3}),\\
&{_2}F_3 \!\left( \frac{5}{6}, \frac{7}{6}; \frac{4}{3}, \frac{5}{3}, \frac{13}{6}; -\frac{4}{9} y^2 \right)= \frac{7 \sqrt{\pi}}{2^{2/3} \!\cdot\! 3^{5/6} \Gamma(5/6)} \, y^{-5/3} +\frac{7 \sqrt{\pi}\, \Gamma(-1/3)}{2^{7/3} \!\cdot\! 3^{13/6} \Gamma(5/6)} \, y^{-7/3}   + O(y^{-3}),\\
&{_2}F_3 \! \left(\! -\frac{1}{6}, \frac{1}{6}; -\frac{1}{3}, \frac{1}{3}, \frac{7}{6}; -\frac{4}{9} y^2 \right)= \frac{\Gamma(4/3)}{3^{5/6}\Gamma(5/3)} \, y^{1/3} +  \frac{\pi}{2^{2/3} \! \cdot \!3^{5/3}\Gamma(5/3)} \, y^{-1/3} + O(y^{-1}), \\
&{_2}F_3 \!\left( \frac{1}{2}, \frac{5}{6}; \frac{1}{3}, \frac{5}{3}, \frac{11}{6}; -\frac{4}{9} y^2 \right)= - \frac{5}{2 \sqrt{3}} \, y^{-1} - \frac{5 \, \Gamma(-1/3)}{2^{5/3} \! \cdot \! 3^{5/6}} \, y^{-5/3} + O(y^{-7/3}),\\
&{_2}F_3 \!\left( \frac{7}{6}, \frac{3}{2}; \frac{5}{3}, \frac{7}{3}, \frac{5}{2}; -\frac{4}{9} y^2 \right)= \frac{2^{2/3} \! \cdot \! 3^{5/6}  \! \cdot \! 6 \sqrt{\pi}}{\Gamma(1/6)} \, y^{-7/3} - \frac{27 \sqrt{3 \pi} \, \Gamma(5/3)}{4 \, \Gamma(1/6)} \, y^{-3} + O(y^{-11/3}),
\end{align}
\end{subequations}
we get
\begin{subequations}
\begin{align}
&\mathcal{I}_1(3)= \frac{\Lambda^2}{4} + \frac{\Gamma(-1/3)}{2^{11/3}\! \cdot \!3^{4/3}} \left( \frac{r_i}{\tilde{\tau}} \right)^{2/3}\!+ O \!\left( \! \frac{r_i^{4/3}}{\Lambda^4 \tilde{\tau}^{4/3}} \! \right),\\
&\mathcal{I}_2(3)= \frac{\Lambda^2}{4} - \frac{\Gamma(-1/3)}{2^{11/3}\! \cdot \!3^{1/3}} \left( \frac{r_i}{\tilde{\tau}} \right)^{2/3}\!+ O \!\left( \! \frac{r_i^{4/3}}{\Lambda^4 \tilde{\tau}^{4/3}} \! \right).
\end{align}
\end{subequations}
Using these results, we can recover Eq.~(\ref{eq:rc_large-3d}).

For $d=4$, we have
\begin{align}
\mathcal{I}_1(4)+\mathcal{I}_2(4)=\Lambda^3 &\left\lbrace \frac{1}{12 \sqrt{3}} \, y^{-1} + \frac{\pi}{12} \left[ 2 y^{1/3}\left( \Ai^2(-y^{2/3}) +  \Bi^2(-y^{2/3}) \right) +2 y^{-1/3}\left( \Ai'^{\,2}(-y^{2/3}) + \Bi'^{\,2}(-y^{2/3}) \right)\right. \right. \nonumber \\
&\left.\left. - y^{-1}\left( \Ai(-y^{2/3})\Ai'(-y^{2/3}) +  \Bi(-y^{2/3})\Bi'(-y^{2/3}) \right) \right] \vphantom{\frac{1}{12 \sqrt{3}}}\! \right\rbrace,
\end{align}
\ew
where we introduced $y=\Lambda^3 \tilde{\tau}/r_i$.
Expanding the Airy functions for large and negative arguments, we finally get
\beq
\mathcal{I}_1(4)+\mathcal{I}_2(4)= -\frac{\Lambda^3}{3} + \frac{1}{12 \sqrt{3}} \left(\frac{r_i}{\tilde{\tau}}\right) +O \! \left( \frac{r_i^2}{\Lambda^6 \tilde{\tau}^2} \right),
\eeq
from which Eq.~(\ref{eq:rc_large-4d}) follows.


\section{Moment generating function}
\label{app:D}

In this section we derive the moment generating function, defined as
\beq
G(s, t)= \bra{\psi(t)} e^{-s\hat{\mathcal{N}}}  \ket{\psi(t)},
\eeq
where $\hat{\mathcal{N}}$ is the operator describing the number of excitations, $\ket{\psi(t)}= U(t)\ket{0}$ is the evolved state at time $t$, and $\ket{0}$ indicates the initial ground state.

Since the effective theory is quadratic and different $k$-modes are coupled only via $r(t)$, the moment generating function can be factorized as
\beq
G(s, t)= \prod_k G^{\phantom{\star}}_k(s, t),
\eeq
where $G^{\phantom{\star}}_k(s, t)$ is the moment generating function for a single $k$-mode.

In order to compute $G^{\phantom{\star}}_k(s, t)$, we have to write the evolved state $\ket{\psi(t)}$ in terms of the operators $\ad_\bk$ and $\ac_\bk$ diagonalizing the initial Hamiltonian.
To this purpose, we introduce a time-dependent operator $\tilde{a}^{\phantom{\dagger}}_\bk(t)$ such that $\tilde{a}^{\phantom{\dagger}}_\bk(t) \ket{\psi(t)}=0$.
Since $\ad_\bk= U^\dagger(t) \tilde{a}^{\phantom{\dagger}}_\bk(t) U(t)$, using Eq.~(\ref{eq:phi}), we have that
\begin{subequations} \label{eq:phi0-pi0}
\begin{align}
&\phi_\bk(0) = \fk (t) \tilde{a}^{\phantom{\dagger}}_\bk(t) + \fks (t) \tilde{a}^\dagger_{-\bk}(t), \\
&\Pi_\bk(0) = \dot{f}^{\phantom{\star}}_k(t) \tilde{a}^{\phantom{\dagger}}_\bk(t) + \dot{f}^\star_k(t) \tilde{a}^\dagger_{-\bk}(t).
\end{align}
\end{subequations}
Furthermore, we know that at $t=0$ 
\begin{subequations} \label{eq:phi0-pi0_2}
\begin{align}
&\phi_\bk(0) = \frac{1}{\sqrt{2 \omega_k(0)}} ( \ad_\bk + \ac_{-\bk} ), \\
&\Pi_\bk(0) = i \sqrt{\frac{\omega_k(0)}{2}} (\ac_{-\bk} - \ad_\bk ) .
\end{align}
\end{subequations}
Combining Eqs.~(\ref{eq:phi0-pi0}) and~(\ref{eq:phi0-pi0_2}), and using the fact that $\fk (t) \dot{f}^\star_k(t) -  \fks (t) \dot{f}^{\phantom{\star}}_k(t) = i$, we get
\beq \label{eq:a_tilda}
\tilde{a}^{\phantom{\dagger}}_\bk(t)= \alpha^\star_k (t) \ad_\bk - \beta^\star_k (t) \ac_{-\bk},
\eeq
with
\begin{subequations}
\begin{align}
&\alpha^{\phantom{\star}}_k(t) = \sqrt{\frac{\omega_k(0)}{2}}  \fk (t) + \frac{i}{\sqrt{2 \omega_k(0)}}  \dot{f}^{\phantom{\star}}_k(t), \\
&\beta^{\phantom{\star}}_k(t) = \sqrt{\frac{\omega_k(0)}{2}}  \fk (t) - \frac{i}{\sqrt{2 \omega_k(0)}}  \dot{f}^{\phantom{\star}}_k(t).
\end{align}
\end{subequations}
Since the evolved state must be annihilated by the operator $\tilde{a}^{\phantom{\dagger}}_\bk(t)$ of Eq.~(\ref{eq:a_tilda}), we finally obtain
\beq
\ket{\psi(t)}_k= \frac{1}{\sqrt{\lvert \alpha^{\phantom{\star}}_k(t) \rvert}} \exp \left( \frac{\beta^\star_k (t)}{2 \alpha^\star_k (t)} \ac_\bk \ac_{-\bk}\right) \ket{0},
\eeq
with $\ad_\bk \ket{0}=0$.

Now we can readily compute the moment generating function for a single $k$-mode, that is
\beq
G^{\phantom{\star}}_k(s, t)= \frac{1}{\sqrt{1+ \rho^{\phantom{\star}}_k(t) \left( 1-e^{-2s} \right)}},
\eeq
where
\beq
\rho^{\phantom{\star}}_k(t)=\lvert \beta^{\phantom{\star}}_k(t) \rvert^2 = \frac{1}{2} \left( \omega_k(0) \lvert \fk (t) \rvert^2 + \frac{\lvert \dot{f}^{\phantom{\star}}_k(t) \rvert^2}{\omega_k(0)} -1 \right).
\eeq

Using the relation
\beq
\ln G(s, t) = L^d \int^{\Lambda} \!\!\! \frac{d^d k}{(2 \pi)^d} \ln G^{\phantom{\star}}_k(s, t),
\eeq
we finally recover the result of Eq.~(\ref{eq:logG}).


\bibliography{Nonequilibrium}

\begin{thebibliography}{44}%
\makeatletter
\providecommand \@ifxundefined [1]{%
 \@ifx{#1\undefined}
}%
\providecommand \@ifnum [1]{%
 \ifnum #1\expandafter \@firstoftwo
 \else \expandafter \@secondoftwo
 \fi
}%
\providecommand \@ifx [1]{%
 \ifx #1\expandafter \@firstoftwo
 \else \expandafter \@secondoftwo
 \fi
}%
\providecommand \natexlab [1]{#1}%
\providecommand \enquote  [1]{``#1''}%
\providecommand \bibnamefont  [1]{#1}%
\providecommand \bibfnamefont [1]{#1}%
\providecommand \citenamefont [1]{#1}%
\providecommand \href@noop [0]{\@secondoftwo}%
\providecommand \href [0]{\begingroup \@sanitize@url \@href}%
\providecommand \@href[1]{\@@startlink{#1}\@@href}%
\providecommand \@@href[1]{\endgroup#1\@@endlink}%
\providecommand \@sanitize@url [0]{\catcode `\\12\catcode `\$12\catcode
  `\&12\catcode `\#12\catcode `\^12\catcode `\_12\catcode `\%12\relax}%
\providecommand \@@startlink[1]{}%
\providecommand \@@endlink[0]{}%
\providecommand \url  [0]{\begingroup\@sanitize@url \@url }%
\providecommand \@url [1]{\endgroup\@href {#1}{\urlprefix }}%
\providecommand \urlprefix  [0]{URL }%
\providecommand \Eprint [0]{\href }%
\providecommand \doibase [0]{http://dx.doi.org/}%
\providecommand \selectlanguage [0]{\@gobble}%
\providecommand \bibinfo  [0]{\@secondoftwo}%
\providecommand \bibfield  [0]{\@secondoftwo}%
\providecommand \translation [1]{[#1]}%
\providecommand \BibitemOpen [0]{}%
\providecommand \bibitemStop [0]{}%
\providecommand \bibitemNoStop [0]{.\EOS\space}%
\providecommand \EOS [0]{\spacefactor3000\relax}%
\providecommand \BibitemShut  [1]{\csname bibitem#1\endcsname}%
\let\auto@bib@innerbib\@empty
\bibitem [{\citenamefont {Polkovnikov}\ \emph {et~al.}(2011)\citenamefont
  {Polkovnikov}, \citenamefont {Sengupta}, \citenamefont {Silva},\ and\
  \citenamefont {Vengalattore}}]{Polkovnikov2011}%
  \BibitemOpen
  \bibfield  {author} {\bibinfo {author} {\bibfnamefont {A.}~\bibnamefont
  {Polkovnikov}}, \bibinfo {author} {\bibfnamefont {K.}~\bibnamefont
  {Sengupta}}, \bibinfo {author} {\bibfnamefont {A.}~\bibnamefont {Silva}}, \
  and\ \bibinfo {author} {\bibfnamefont {M.}~\bibnamefont {Vengalattore}},\
  }\href {\doibase 10.1103/RevModPhys.83.863} {\bibfield  {journal} {\bibinfo
  {journal} {Rev. Mod. Phys.}\ }\textbf {\bibinfo {volume} {83}},\ \bibinfo
  {pages} {863} (\bibinfo {year} {2011})}\BibitemShut {NoStop}%
\bibitem [{\citenamefont {Lamacraft}\ and\ \citenamefont
  {Moore}(2012)}]{Lamacraft2012}%
  \BibitemOpen
  \bibfield  {author} {\bibinfo {author} {\bibfnamefont {A.}~\bibnamefont
  {Lamacraft}}\ and\ \bibinfo {author} {\bibfnamefont {J.}~\bibnamefont
  {Moore}},\ }in\ \href {\doibase 10.1016/B978-0-444-53857-4.00007-6} {\emph
  {\bibinfo {booktitle} {Ultracold Bosonic and Fermionic Gases}}},\ \bibinfo
  {series} {Contemporary Concepts of Condensed Matter Science}, Vol.~\bibinfo
  {volume} {5},\ \bibinfo {editor} {edited by\ \bibinfo {editor} {\bibfnamefont
  {A.~L.~F.}\ \bibnamefont {Kathryn~Levin}}\ and\ \bibinfo {editor}
  {\bibfnamefont {D.~M.}\ \bibnamefont {Stamper-Kurn}}}\ (\bibinfo  {publisher}
  {Elsevier},\ \bibinfo {year} {2012})\ pp.\ \bibinfo {pages} {177 --
  202}\BibitemShut {NoStop}%
\bibitem [{\citenamefont {Dziarmaga}(2010)}]{Dziarmaga2010}%
  \BibitemOpen
  \bibfield  {author} {\bibinfo {author} {\bibfnamefont {J.}~\bibnamefont
  {Dziarmaga}},\ }\href {\doibase 10.1080/00018732.2010.514702} {\bibfield
  {journal} {\bibinfo  {journal} {Adv. in Phys.}\ }\textbf {\bibinfo {volume}
  {59}},\ \bibinfo {pages} {1063} (\bibinfo {year} {2010})}\BibitemShut
  {NoStop}%
\bibitem [{\citenamefont {Yukalov}(2011)}]{yukalov2011}%
  \BibitemOpen
  \bibfield  {author} {\bibinfo {author} {\bibfnamefont {V.~I.}\ \bibnamefont
  {Yukalov}},\ }\href@noop {} {\bibfield  {journal} {\bibinfo  {journal} {Laser
  Physics Letters}\ }\textbf {\bibinfo {volume} {8}},\ \bibinfo {pages} {485}
  (\bibinfo {year} {2011})}\BibitemShut {NoStop}%
\bibitem [{\citenamefont {Bloch}\ \emph {et~al.}(2008)\citenamefont {Bloch},
  \citenamefont {Dalibard},\ and\ \citenamefont {Zwerger}}]{bloch_review}%
  \BibitemOpen
  \bibfield  {author} {\bibinfo {author} {\bibfnamefont {I.}~\bibnamefont
  {Bloch}}, \bibinfo {author} {\bibfnamefont {J.}~\bibnamefont {Dalibard}}, \
  and\ \bibinfo {author} {\bibfnamefont {W.}~\bibnamefont {Zwerger}},\
  }\href@noop {} {\bibfield  {journal} {\bibinfo  {journal} {Rev. Mod. Phys.}\
  }\textbf {\bibinfo {volume} {80}},\ \bibinfo {pages} {885} (\bibinfo {year}
  {2008})}\BibitemShut {NoStop}%
\bibitem [{\citenamefont {Greiner}\ \emph
  {et~al.}(2002{\natexlab{a}})\citenamefont {Greiner}, \citenamefont {Mandel},
  \citenamefont {Esslinger}, \citenamefont {H\"{a}nsch},\ and\ \citenamefont
  {Bloch}}]{greiner2002a}%
  \BibitemOpen
  \bibfield  {author} {\bibinfo {author} {\bibfnamefont {M.}~\bibnamefont
  {Greiner}}, \bibinfo {author} {\bibfnamefont {O.}~\bibnamefont {Mandel}},
  \bibinfo {author} {\bibfnamefont {T.}~\bibnamefont {Esslinger}}, \bibinfo
  {author} {\bibfnamefont {T.~W.}\ \bibnamefont {H\"{a}nsch}}, \ and\ \bibinfo
  {author} {\bibfnamefont {I.}~\bibnamefont {Bloch}},\ }\href@noop {}
  {\bibfield  {journal} {\bibinfo  {journal} {Nature}\ }\textbf {\bibinfo
  {volume} {415}},\ \bibinfo {pages} {39} (\bibinfo {year}
  {2002}{\natexlab{a}})}\BibitemShut {NoStop}%
\bibitem [{\citenamefont {Greiner}\ \emph
  {et~al.}(2002{\natexlab{b}})\citenamefont {Greiner}, \citenamefont {Mandel},
  \citenamefont {Hansch},\ and\ \citenamefont {Bloch}}]{greiner2002b}%
  \BibitemOpen
  \bibfield  {author} {\bibinfo {author} {\bibfnamefont {M.}~\bibnamefont
  {Greiner}}, \bibinfo {author} {\bibfnamefont {O.}~\bibnamefont {Mandel}},
  \bibinfo {author} {\bibfnamefont {T.~W.}\ \bibnamefont {Hansch}}, \ and\
  \bibinfo {author} {\bibfnamefont {I.}~\bibnamefont {Bloch}},\ }\href
  {http://dx.doi.org/10.1038/nature00968} {\bibfield  {journal} {\bibinfo
  {journal} {Nature}\ }\textbf {\bibinfo {volume} {419}},\ \bibinfo {pages}
  {51} (\bibinfo {year} {2002}{\natexlab{b}})}\BibitemShut {NoStop}%
\bibitem [{\citenamefont {Sadler}\ \emph {et~al.}(2006)\citenamefont {Sadler},
  \citenamefont {Higbie}, \citenamefont {Leslie}, \citenamefont
  {Vengalattore},\ and\ \citenamefont {Stamper-Kurn}}]{sadler_06}%
  \BibitemOpen
  \bibfield  {author} {\bibinfo {author} {\bibfnamefont {L.~E.}\ \bibnamefont
  {Sadler}}, \bibinfo {author} {\bibfnamefont {J.~M.}\ \bibnamefont {Higbie}},
  \bibinfo {author} {\bibfnamefont {S.~R.}\ \bibnamefont {Leslie}}, \bibinfo
  {author} {\bibfnamefont {M.}~\bibnamefont {Vengalattore}}, \ and\ \bibinfo
  {author} {\bibfnamefont {D.~M.}\ \bibnamefont {Stamper-Kurn}},\ }\href@noop
  {} {\bibfield  {journal} {\bibinfo  {journal} {Nature}\ }\textbf {\bibinfo
  {volume} {443}},\ \bibinfo {pages} {312} (\bibinfo {year}
  {2006})}\BibitemShut {NoStop}%
\bibitem [{\citenamefont {Kinoshita}\ \emph {et~al.}(2006)\citenamefont
  {Kinoshita}, \citenamefont {Wenger},\ and\ \citenamefont
  {Weiss}}]{kinoshita}%
  \BibitemOpen
  \bibfield  {author} {\bibinfo {author} {\bibfnamefont {T.}~\bibnamefont
  {Kinoshita}}, \bibinfo {author} {\bibfnamefont {T.}~\bibnamefont {Wenger}}, \
  and\ \bibinfo {author} {\bibfnamefont {D.~S.}\ \bibnamefont {Weiss}},\ }\href
  {http://dx.doi.org/10.1038/nature04693} {\bibfield  {journal} {\bibinfo
  {journal} {Nature}\ }\textbf {\bibinfo {volume} {440}},\ \bibinfo {pages}
  {900} (\bibinfo {year} {2006})}\BibitemShut {NoStop}%
\bibitem [{\citenamefont {Kitagawa}\ \emph {et~al.}(2011)\citenamefont
  {Kitagawa}, \citenamefont {Imambekov}, \citenamefont {Schmiedmayer},\ and\
  \citenamefont {Demler}}]{Kitagawa2011}%
  \BibitemOpen
  \bibfield  {author} {\bibinfo {author} {\bibfnamefont {T.}~\bibnamefont
  {Kitagawa}}, \bibinfo {author} {\bibfnamefont {A.}~\bibnamefont {Imambekov}},
  \bibinfo {author} {\bibfnamefont {J.}~\bibnamefont {Schmiedmayer}}, \ and\
  \bibinfo {author} {\bibfnamefont {E.}~\bibnamefont {Demler}},\ }\href
  {http://stacks.iop.org/1367-2630/13/i=7/a=073018} {\bibfield  {journal}
  {\bibinfo  {journal} {New J. Phys.}\ }\textbf {\bibinfo {volume} {13}},\
  \bibinfo {pages} {073018} (\bibinfo {year} {2011})}\BibitemShut {NoStop}%
\bibitem [{\citenamefont {Gring}\ \emph {et~al.}(2012)\citenamefont {Gring},
  \citenamefont {Kuhnert}, \citenamefont {Langen}, \citenamefont {Kitagawa},
  \citenamefont {Rauer}, \citenamefont {Schreitl}, \citenamefont {Mazets},
  \citenamefont {Smith}, \citenamefont {Demler},\ and\ \citenamefont
  {Schmiedmayer}}]{Gring2012}%
  \BibitemOpen
  \bibfield  {author} {\bibinfo {author} {\bibfnamefont {M.}~\bibnamefont
  {Gring}}, \bibinfo {author} {\bibfnamefont {M.}~\bibnamefont {Kuhnert}},
  \bibinfo {author} {\bibfnamefont {T.}~\bibnamefont {Langen}}, \bibinfo
  {author} {\bibfnamefont {T.}~\bibnamefont {Kitagawa}}, \bibinfo {author}
  {\bibfnamefont {B.}~\bibnamefont {Rauer}}, \bibinfo {author} {\bibfnamefont
  {M.}~\bibnamefont {Schreitl}}, \bibinfo {author} {\bibfnamefont
  {I.}~\bibnamefont {Mazets}}, \bibinfo {author} {\bibfnamefont {D.~A.}\
  \bibnamefont {Smith}}, \bibinfo {author} {\bibfnamefont {E.}~\bibnamefont
  {Demler}}, \ and\ \bibinfo {author} {\bibfnamefont {J.}~\bibnamefont
  {Schmiedmayer}},\ }\href {\doibase 10.1126/science.1224953} {\bibfield
  {journal} {\bibinfo  {journal} {Science}\ }\textbf {\bibinfo {volume}
  {337}},\ \bibinfo {pages} {1318} (\bibinfo {year} {2012})}\BibitemShut
  {NoStop}%
\bibitem [{\citenamefont {Langen}\ \emph {et~al.}(2013)\citenamefont {Langen},
  \citenamefont {Geiger}, \citenamefont {Kuhnert}, \citenamefont {Rauer},\ and\
  \citenamefont {Schmiedmayer}}]{langen2013}%
  \BibitemOpen
  \bibfield  {author} {\bibinfo {author} {\bibfnamefont {T.}~\bibnamefont
  {Langen}}, \bibinfo {author} {\bibfnamefont {R.}~\bibnamefont {Geiger}},
  \bibinfo {author} {\bibfnamefont {M.}~\bibnamefont {Kuhnert}}, \bibinfo
  {author} {\bibfnamefont {B.}~\bibnamefont {Rauer}}, \ and\ \bibinfo {author}
  {\bibfnamefont {J.}~\bibnamefont {Schmiedmayer}},\ }\href@noop {} {\bibfield
  {journal} {\bibinfo  {journal} {Nature Physics}\ }\textbf {\bibinfo {volume}
  {9}},\ \bibinfo {pages} {640} (\bibinfo {year} {2013})}\BibitemShut {NoStop}%
\bibitem [{\citenamefont {Cheneau}\ \emph {et~al.}(2012)\citenamefont
  {Cheneau}, \citenamefont {Barmettler}, \citenamefont {Poletti}, \citenamefont
  {Endres}, \citenamefont {Schaub}, \citenamefont {Fukuhara}, \citenamefont
  {Gross}, \citenamefont {Bloch}, \citenamefont {Kollath},\ and\ \citenamefont
  {Kuhr}}]{cheneau_2012}%
  \BibitemOpen
  \bibfield  {author} {\bibinfo {author} {\bibfnamefont {M.}~\bibnamefont
  {Cheneau}}, \bibinfo {author} {\bibfnamefont {P.}~\bibnamefont {Barmettler}},
  \bibinfo {author} {\bibfnamefont {D.}~\bibnamefont {Poletti}}, \bibinfo
  {author} {\bibfnamefont {M.}~\bibnamefont {Endres}}, \bibinfo {author}
  {\bibfnamefont {P.}~\bibnamefont {Schaub}}, \bibinfo {author} {\bibfnamefont
  {T.}~\bibnamefont {Fukuhara}}, \bibinfo {author} {\bibfnamefont
  {C.}~\bibnamefont {Gross}}, \bibinfo {author} {\bibfnamefont
  {I.}~\bibnamefont {Bloch}}, \bibinfo {author} {\bibfnamefont
  {C.}~\bibnamefont {Kollath}}, \ and\ \bibinfo {author} {\bibfnamefont
  {S.}~\bibnamefont {Kuhr}},\ }\href@noop {} {\bibfield  {journal} {\bibinfo
  {journal} {Nature}\ }\textbf {\bibinfo {volume} {481}},\ \bibinfo {pages}
  {484} (\bibinfo {year} {2012})}\BibitemShut {NoStop}%
\bibitem [{\citenamefont {Deutsch}(1991)}]{deutsch_91}%
  \BibitemOpen
  \bibfield  {author} {\bibinfo {author} {\bibfnamefont {J.}~\bibnamefont
  {Deutsch}},\ }\href@noop {} {\bibfield  {journal} {\bibinfo  {journal} {Phys.
  Rev. A}\ }\textbf {\bibinfo {volume} {43}},\ \bibinfo {pages} {2046}
  (\bibinfo {year} {1991})}\BibitemShut {NoStop}%
\bibitem [{\citenamefont {Srednicki}(1994)}]{srednicki_94}%
  \BibitemOpen
  \bibfield  {author} {\bibinfo {author} {\bibfnamefont {M.}~\bibnamefont
  {Srednicki}},\ }\href@noop {} {\bibfield  {journal} {\bibinfo  {journal}
  {Phys. Rev. E}\ }\textbf {\bibinfo {volume} {50}},\ \bibinfo {pages} {888}
  (\bibinfo {year} {1994})}\BibitemShut {NoStop}%
\bibitem [{\citenamefont {Rigol}\ \emph {et~al.}(2008)\citenamefont {Rigol},
  \citenamefont {Dunjko},\ and\ \citenamefont {Olshanii}}]{rigol_08}%
  \BibitemOpen
  \bibfield  {author} {\bibinfo {author} {\bibfnamefont {M.}~\bibnamefont
  {Rigol}}, \bibinfo {author} {\bibfnamefont {V.}~\bibnamefont {Dunjko}}, \
  and\ \bibinfo {author} {\bibfnamefont {M.}~\bibnamefont {Olshanii}},\
  }\href@noop {} {\bibfield  {journal} {\bibinfo  {journal} {Nature}\ }\textbf
  {\bibinfo {volume} {452}},\ \bibinfo {pages} {854} (\bibinfo {year}
  {2008})}\BibitemShut {NoStop}%
\bibitem [{\citenamefont {Rigol}\ \emph {et~al.}(2007)\citenamefont {Rigol},
  \citenamefont {Dunjko}, \citenamefont {Yurovsky},\ and\ \citenamefont
  {Olshanii}}]{rigol_07}%
  \BibitemOpen
  \bibfield  {author} {\bibinfo {author} {\bibfnamefont {M.}~\bibnamefont
  {Rigol}}, \bibinfo {author} {\bibfnamefont {V.}~\bibnamefont {Dunjko}},
  \bibinfo {author} {\bibfnamefont {V.}~\bibnamefont {Yurovsky}}, \ and\
  \bibinfo {author} {\bibfnamefont {M.}~\bibnamefont {Olshanii}},\ }\href@noop
  {} {\bibfield  {journal} {\bibinfo  {journal} {Phys. Rev. Lett.}\ }\textbf
  {\bibinfo {volume} {98}},\ \bibinfo {pages} {050405} (\bibinfo {year}
  {2007})}\BibitemShut {NoStop}%
\bibitem [{\citenamefont {Barthel}\ and\ \citenamefont
  {Schollw\"ock}(2008)}]{barthel_08}%
  \BibitemOpen
  \bibfield  {author} {\bibinfo {author} {\bibfnamefont {T.}~\bibnamefont
  {Barthel}}\ and\ \bibinfo {author} {\bibfnamefont {U.}~\bibnamefont
  {Schollw\"ock}},\ }\href@noop {} {\bibfield  {journal} {\bibinfo  {journal}
  {Phys. Rev. Lett.}\ }\textbf {\bibinfo {volume} {100}},\ \bibinfo {pages}
  {100601} (\bibinfo {year} {2008})}\BibitemShut {NoStop}%
\bibitem [{\citenamefont {Kollar}\ and\ \citenamefont
  {Eckstein}(2008)}]{kollar_08}%
  \BibitemOpen
  \bibfield  {author} {\bibinfo {author} {\bibfnamefont {M.}~\bibnamefont
  {Kollar}}\ and\ \bibinfo {author} {\bibfnamefont {M.}~\bibnamefont
  {Eckstein}},\ }\href@noop {} {\bibfield  {journal} {\bibinfo  {journal}
  {Phys. Rev. A}\ }\textbf {\bibinfo {volume} {78}},\ \bibinfo {pages} {013626}
  (\bibinfo {year} {2008})}\BibitemShut {NoStop}%
\bibitem [{\citenamefont {Iucci}\ and\ \citenamefont
  {Cazalilla}(2009)}]{iucci_09}%
  \BibitemOpen
  \bibfield  {author} {\bibinfo {author} {\bibfnamefont {A.}~\bibnamefont
  {Iucci}}\ and\ \bibinfo {author} {\bibfnamefont {M.~A.}\ \bibnamefont
  {Cazalilla}},\ }\href {\doibase 10.1103/PhysRevA.80.063619} {\bibfield
  {journal} {\bibinfo  {journal} {Phys. Rev. A}\ }\textbf {\bibinfo {volume}
  {80}},\ \bibinfo {pages} {063619} (\bibinfo {year} {2009})}\BibitemShut
  {NoStop}%
\bibitem [{\citenamefont {Cazalilla}\ \emph {et~al.}(2012)\citenamefont
  {Cazalilla}, \citenamefont {Iucci},\ and\ \citenamefont
  {Chung}}]{cazalilla2012}%
  \BibitemOpen
  \bibfield  {author} {\bibinfo {author} {\bibfnamefont {M.~A.}\ \bibnamefont
  {Cazalilla}}, \bibinfo {author} {\bibfnamefont {A.}~\bibnamefont {Iucci}}, \
  and\ \bibinfo {author} {\bibfnamefont {M.-C.}\ \bibnamefont {Chung}},\ }\href
  {\doibase 10.1103/PhysRevE.85.011133} {\bibfield  {journal} {\bibinfo
  {journal} {Phys. Rev. E}\ }\textbf {\bibinfo {volume} {85}},\ \bibinfo
  {pages} {011133} (\bibinfo {year} {2012})}\BibitemShut {NoStop}%
\bibitem [{\citenamefont {Jaynes}(1957)}]{Jaynes1957}%
  \BibitemOpen
  \bibfield  {author} {\bibinfo {author} {\bibfnamefont {E.~T.}\ \bibnamefont
  {Jaynes}},\ }\href {\doibase 10.1103/PhysRev.106.620} {\bibfield  {journal}
  {\bibinfo  {journal} {Phys. Rev.}\ }\textbf {\bibinfo {volume} {106}},\
  \bibinfo {pages} {620} (\bibinfo {year} {1957})}\BibitemShut {NoStop}%
\bibitem [{\citenamefont {Berges}\ \emph {et~al.}(2004)\citenamefont {Berges},
  \citenamefont {Bors\'anyi},\ and\ \citenamefont {Wetterich}}]{Berges2004}%
  \BibitemOpen
  \bibfield  {author} {\bibinfo {author} {\bibfnamefont {J.}~\bibnamefont
  {Berges}}, \bibinfo {author} {\bibfnamefont {S.}~\bibnamefont {Bors\'anyi}},
  \ and\ \bibinfo {author} {\bibfnamefont {C.}~\bibnamefont {Wetterich}},\
  }\href {\doibase 10.1103/PhysRevLett.93.142002} {\bibfield  {journal}
  {\bibinfo  {journal} {Phys. Rev. Lett.}\ }\textbf {\bibinfo {volume} {93}},\
  \bibinfo {pages} {142002} (\bibinfo {year} {2004})}\BibitemShut {NoStop}%
\bibitem [{\citenamefont {Moeckel}\ and\ \citenamefont
  {Kehrein}(2008)}]{moeckel_08}%
  \BibitemOpen
  \bibfield  {author} {\bibinfo {author} {\bibfnamefont {M.}~\bibnamefont
  {Moeckel}}\ and\ \bibinfo {author} {\bibfnamefont {S.}~\bibnamefont
  {Kehrein}},\ }\href@noop {} {\bibfield  {journal} {\bibinfo  {journal} {Phys.
  Rev. Lett.}\ }\textbf {\bibinfo {volume} {100}},\ \bibinfo {pages} {175702}
  (\bibinfo {year} {2008})}\BibitemShut {NoStop}%
\bibitem [{\citenamefont {Moeckel}\ and\ \citenamefont
  {Kehrein}(2009)}]{moeckel_09}%
  \BibitemOpen
  \bibfield  {author} {\bibinfo {author} {\bibfnamefont {M.}~\bibnamefont
  {Moeckel}}\ and\ \bibinfo {author} {\bibfnamefont {S.}~\bibnamefont
  {Kehrein}},\ }\href@noop {} {\bibfield  {journal} {\bibinfo  {journal} {Ann.
  Phys.}\ }\textbf {\bibinfo {volume} {324}},\ \bibinfo {pages} {2146}
  (\bibinfo {year} {2009})}\BibitemShut {NoStop}%
\bibitem [{\citenamefont {Moeckel}\ and\ \citenamefont
  {Kehrein}(2010)}]{moeckel_10}%
  \BibitemOpen
  \bibfield  {author} {\bibinfo {author} {\bibfnamefont {M.}~\bibnamefont
  {Moeckel}}\ and\ \bibinfo {author} {\bibfnamefont {S.}~\bibnamefont
  {Kehrein}},\ }\href@noop {} {\bibfield  {journal} {\bibinfo  {journal} {New
  J. Phys.}\ }\textbf {\bibinfo {volume} {12}},\ \bibinfo {pages} {055016}
  (\bibinfo {year} {2010})}\BibitemShut {NoStop}%
\bibitem [{\citenamefont {Kollar}\ \emph {et~al.}(2011)\citenamefont {Kollar},
  \citenamefont {Wolf},\ and\ \citenamefont {Eckstein}}]{Kollar2011}%
  \BibitemOpen
  \bibfield  {author} {\bibinfo {author} {\bibfnamefont {M.}~\bibnamefont
  {Kollar}}, \bibinfo {author} {\bibfnamefont {F.~A.}\ \bibnamefont {Wolf}}, \
  and\ \bibinfo {author} {\bibfnamefont {M.}~\bibnamefont {Eckstein}},\ }\href
  {\doibase 10.1103/PhysRevB.84.054304} {\bibfield  {journal} {\bibinfo
  {journal} {Phys. Rev. B}\ }\textbf {\bibinfo {volume} {84}},\ \bibinfo
  {pages} {054304} (\bibinfo {year} {2011})}\BibitemShut {NoStop}%
\bibitem [{\citenamefont {Marino}\ and\ \citenamefont
  {Silva}(2012)}]{Marino2012}%
  \BibitemOpen
  \bibfield  {author} {\bibinfo {author} {\bibfnamefont {J.}~\bibnamefont
  {Marino}}\ and\ \bibinfo {author} {\bibfnamefont {A.}~\bibnamefont {Silva}},\
  }\href {\doibase 10.1103/PhysRevB.86.060408} {\bibfield  {journal} {\bibinfo
  {journal} {Phys. Rev. B}\ }\textbf {\bibinfo {volume} {86}},\ \bibinfo
  {pages} {060408} (\bibinfo {year} {2012})}\BibitemShut {NoStop}%
\bibitem [{\citenamefont {Marcuzzi}\ \emph {et~al.}(2013)\citenamefont
  {Marcuzzi}, \citenamefont {Marino}, \citenamefont {Gambassi},\ and\
  \citenamefont {Silva}}]{Marcuzzi2013}%
  \BibitemOpen
  \bibfield  {author} {\bibinfo {author} {\bibfnamefont {M.}~\bibnamefont
  {Marcuzzi}}, \bibinfo {author} {\bibfnamefont {J.}~\bibnamefont {Marino}},
  \bibinfo {author} {\bibfnamefont {A.}~\bibnamefont {Gambassi}}, \ and\
  \bibinfo {author} {\bibfnamefont {A.}~\bibnamefont {Silva}},\ }\href
  {\doibase 10.1103/PhysRevLett.111.197203} {\bibfield  {journal} {\bibinfo
  {journal} {Phys. Rev. Lett.}\ }\textbf {\bibinfo {volume} {111}},\ \bibinfo
  {pages} {197203} (\bibinfo {year} {2013})}\BibitemShut {NoStop}%
\bibitem [{\citenamefont {Essler}\ \emph {et~al.}(2014)\citenamefont {Essler},
  \citenamefont {Kehrein}, \citenamefont {Manmana},\ and\ \citenamefont
  {Robinson}}]{Essler2014}%
  \BibitemOpen
  \bibfield  {author} {\bibinfo {author} {\bibfnamefont {F.~H.~L.}\
  \bibnamefont {Essler}}, \bibinfo {author} {\bibfnamefont {S.}~\bibnamefont
  {Kehrein}}, \bibinfo {author} {\bibfnamefont {S.~R.}\ \bibnamefont
  {Manmana}}, \ and\ \bibinfo {author} {\bibfnamefont {N.~J.}\ \bibnamefont
  {Robinson}},\ }\href {\doibase 10.1103/PhysRevB.89.165104} {\bibfield
  {journal} {\bibinfo  {journal} {Phys. Rev. B}\ }\textbf {\bibinfo {volume}
  {89}},\ \bibinfo {pages} {165104} (\bibinfo {year} {2014})}\BibitemShut
  {NoStop}%
\bibitem [{\citenamefont {Eckstein}\ \emph {et~al.}(2009)\citenamefont
  {Eckstein}, \citenamefont {Kollar},\ and\ \citenamefont
  {Werner}}]{eckstein_2009}%
  \BibitemOpen
  \bibfield  {author} {\bibinfo {author} {\bibfnamefont {M.}~\bibnamefont
  {Eckstein}}, \bibinfo {author} {\bibfnamefont {M.}~\bibnamefont {Kollar}}, \
  and\ \bibinfo {author} {\bibfnamefont {P.}~\bibnamefont {Werner}},\
  }\href@noop {} {\bibfield  {journal} {\bibinfo  {journal} {Phys. Rev. Lett.}\
  }\textbf {\bibinfo {volume} {103}},\ \bibinfo {pages} {056403} (\bibinfo
  {year} {2009})}\BibitemShut {NoStop}%
\bibitem [{\citenamefont {Schir\'o}\ and\ \citenamefont
  {Fabrizio}(2010)}]{schiro_2010}%
  \BibitemOpen
  \bibfield  {author} {\bibinfo {author} {\bibfnamefont {M.}~\bibnamefont
  {Schir\'o}}\ and\ \bibinfo {author} {\bibfnamefont {M.}~\bibnamefont
  {Fabrizio}},\ }\href@noop {} {\bibfield  {journal} {\bibinfo  {journal}
  {Phys. Rev. Lett.}\ }\textbf {\bibinfo {volume} {105}},\ \bibinfo {pages}
  {076401} (\bibinfo {year} {2010})}\BibitemShut {NoStop}%
\bibitem [{\citenamefont {Tsuji}\ \emph {et~al.}(2013)\citenamefont {Tsuji},
  \citenamefont {Eckstein},\ and\ \citenamefont {Werner}}]{Tsuji2013}%
  \BibitemOpen
  \bibfield  {author} {\bibinfo {author} {\bibfnamefont {N.}~\bibnamefont
  {Tsuji}}, \bibinfo {author} {\bibfnamefont {M.}~\bibnamefont {Eckstein}}, \
  and\ \bibinfo {author} {\bibfnamefont {P.}~\bibnamefont {Werner}},\ }\href
  {\doibase 10.1103/PhysRevLett.110.136404} {\bibfield  {journal} {\bibinfo
  {journal} {Phys. Rev. Lett.}\ }\textbf {\bibinfo {volume} {110}},\ \bibinfo
  {pages} {136404} (\bibinfo {year} {2013})}\BibitemShut {NoStop}%
\bibitem [{\citenamefont {Sciolla}\ and\ \citenamefont
  {Biroli}(2010)}]{sciolla_2010}%
  \BibitemOpen
  \bibfield  {author} {\bibinfo {author} {\bibfnamefont {B.}~\bibnamefont
  {Sciolla}}\ and\ \bibinfo {author} {\bibfnamefont {G.}~\bibnamefont
  {Biroli}},\ }\href {\doibase 10.1103/PhysRevLett.105.220401} {\bibfield
  {journal} {\bibinfo  {journal} {Phys. Rev. Lett.}\ }\textbf {\bibinfo
  {volume} {105}},\ \bibinfo {pages} {220401} (\bibinfo {year}
  {2010})}\BibitemShut {NoStop}%
\bibitem [{\citenamefont {Sciolla}\ and\ \citenamefont
  {Biroli}(2011)}]{sciolla_2011}%
  \BibitemOpen
  \bibfield  {author} {\bibinfo {author} {\bibfnamefont {B.}~\bibnamefont
  {Sciolla}}\ and\ \bibinfo {author} {\bibfnamefont {G.}~\bibnamefont
  {Biroli}},\ }\href {http://stacks.iop.org/1742-5468/2011/i=11/a=P11003}
  {\bibfield  {journal} {\bibinfo  {journal} {J. Stat. Mech.}\ }\textbf
  {\bibinfo {volume} {2011}},\ \bibinfo {pages} {P11003} (\bibinfo {year}
  {2011})}\BibitemShut {NoStop}%
\bibitem [{\citenamefont {Gambassi}\ and\ \citenamefont
  {Calabrese}(2011)}]{Gambassi2011a}%
  \BibitemOpen
  \bibfield  {author} {\bibinfo {author} {\bibfnamefont {A.}~\bibnamefont
  {Gambassi}}\ and\ \bibinfo {author} {\bibfnamefont {P.}~\bibnamefont
  {Calabrese}},\ }\href {\doibase 10.1209/0295-5075/95/66007} {\bibfield
  {journal} {\bibinfo  {journal} {Europhys. Lett.}\ }\textbf {\bibinfo {volume}
  {95}},\ \bibinfo {pages} {66007} (\bibinfo {year} {2011})}\BibitemShut
  {NoStop}%
\bibitem [{\citenamefont {Sciolla}\ and\ \citenamefont
  {Biroli}(2013)}]{sciolla_2013}%
  \BibitemOpen
  \bibfield  {author} {\bibinfo {author} {\bibfnamefont {B.}~\bibnamefont
  {Sciolla}}\ and\ \bibinfo {author} {\bibfnamefont {G.}~\bibnamefont
  {Biroli}},\ }\href {\doibase 10.1103/PhysRevB.88.201110} {\bibfield
  {journal} {\bibinfo  {journal} {Phys. Rev. B}\ }\textbf {\bibinfo {volume}
  {88}},\ \bibinfo {pages} {201110} (\bibinfo {year} {2013})}\BibitemShut
  {NoStop}%
\bibitem [{\citenamefont {Smacchia}\ \emph {et~al.}(2015)\citenamefont
  {Smacchia}, \citenamefont {Knap}, \citenamefont {Demler},\ and\ \citenamefont
  {Silva}}]{Smacchia2015}%
  \BibitemOpen
  \bibfield  {author} {\bibinfo {author} {\bibfnamefont {P.}~\bibnamefont
  {Smacchia}}, \bibinfo {author} {\bibfnamefont {M.}~\bibnamefont {Knap}},
  \bibinfo {author} {\bibfnamefont {E.}~\bibnamefont {Demler}}, \ and\ \bibinfo
  {author} {\bibfnamefont {A.}~\bibnamefont {Silva}},\ }\href {\doibase
  10.1103/PhysRevB.91.205136} {\bibfield  {journal} {\bibinfo  {journal} {Phys.
  Rev. B}\ }\textbf {\bibinfo {volume} {91}},\ \bibinfo {pages} {205136}
  (\bibinfo {year} {2015})}\BibitemShut {NoStop}%
\bibitem [{\citenamefont {Chiocchetta}\ \emph {et~al.}(2015)\citenamefont
  {Chiocchetta}, \citenamefont {Tavora}, \citenamefont {Gambassi},\ and\
  \citenamefont {Mitra}}]{Chiocchetta2015}%
  \BibitemOpen
  \bibfield  {author} {\bibinfo {author} {\bibfnamefont {A.}~\bibnamefont
  {Chiocchetta}}, \bibinfo {author} {\bibfnamefont {M.}~\bibnamefont {Tavora}},
  \bibinfo {author} {\bibfnamefont {A.}~\bibnamefont {Gambassi}}, \ and\
  \bibinfo {author} {\bibfnamefont {A.}~\bibnamefont {Mitra}},\ }\href
  {\doibase 10.1103/PhysRevB.91.220302} {\bibfield  {journal} {\bibinfo
  {journal} {Phys. Rev. B}\ }\textbf {\bibinfo {volume} {91}},\ \bibinfo
  {pages} {220302} (\bibinfo {year} {2015})}\BibitemShut {NoStop}%
\bibitem [{\citenamefont {Moshe}\ and\ \citenamefont
  {Zinn-Justin}(2003)}]{moshe_2003}%
  \BibitemOpen
  \bibfield  {author} {\bibinfo {author} {\bibfnamefont {M.}~\bibnamefont
  {Moshe}}\ and\ \bibinfo {author} {\bibfnamefont {J.}~\bibnamefont
  {Zinn-Justin}},\ }\href@noop {} {\bibfield  {journal} {\bibinfo  {journal}
  {Phys. Rep.}\ }\textbf {\bibinfo {volume} {385}},\ \bibinfo {pages} {69}
  (\bibinfo {year} {2003})}\BibitemShut {NoStop}%
\bibitem [{\citenamefont {Sotiriadis}\ and\ \citenamefont
  {Cardy}(2010)}]{SC10}%
  \BibitemOpen
  \bibfield  {author} {\bibinfo {author} {\bibfnamefont {S.}~\bibnamefont
  {Sotiriadis}}\ and\ \bibinfo {author} {\bibfnamefont {J.}~\bibnamefont
  {Cardy}},\ }\href {\doibase 10.1103/PhysRevB.81.134305} {\bibfield  {journal}
  {\bibinfo  {journal} {Phys. Rev. B}\ }\textbf {\bibinfo {volume} {81}},\
  \bibinfo {pages} {134305} (\bibinfo {year} {2010})}\BibitemShut {NoStop}%
\bibitem [{\citenamefont {Chandran}\ \emph {et~al.}(2013)\citenamefont
  {Chandran}, \citenamefont {Nanduri}, \citenamefont {Gubser},\ and\
  \citenamefont {Sondhi}}]{Chandran2013}%
  \BibitemOpen
  \bibfield  {author} {\bibinfo {author} {\bibfnamefont {A.}~\bibnamefont
  {Chandran}}, \bibinfo {author} {\bibfnamefont {A.}~\bibnamefont {Nanduri}},
  \bibinfo {author} {\bibfnamefont {S.~S.}\ \bibnamefont {Gubser}}, \ and\
  \bibinfo {author} {\bibfnamefont {S.~L.}\ \bibnamefont {Sondhi}},\ }\href
  {\doibase 10.1103/PhysRevB.88.024306} {\bibfield  {journal} {\bibinfo
  {journal} {Phys. Rev. B}\ }\textbf {\bibinfo {volume} {88}},\ \bibinfo
  {pages} {024306} (\bibinfo {year} {2013})}\BibitemShut {NoStop}%
\bibitem [{\citenamefont {Calabrese}\ and\ \citenamefont
  {Cardy}(2007)}]{calabrese_07}%
  \BibitemOpen
  \bibfield  {author} {\bibinfo {author} {\bibfnamefont {P.}~\bibnamefont
  {Calabrese}}\ and\ \bibinfo {author} {\bibfnamefont {J.}~\bibnamefont
  {Cardy}},\ }\href {http://stacks.iop.org/1742-5468/2007/i=06/a=P06008}
  {\bibfield  {journal} {\bibinfo  {journal} {J. Stat. Mech: Th. and Exp.}\
  }\textbf {\bibinfo {volume} {2007}},\ \bibinfo {pages} {P06008} (\bibinfo
  {year} {2007})}\BibitemShut {NoStop}%
\bibitem [{\citenamefont {Abramowitz}\ and\ \citenamefont
  {Stegun}(1964)}]{abramowitz}%
  \BibitemOpen
  \bibfield  {author} {\bibinfo {author} {\bibfnamefont {M.}~\bibnamefont
  {Abramowitz}}\ and\ \bibinfo {author} {\bibfnamefont {I.}~\bibnamefont
  {Stegun}},\ }\href@noop {} {\emph {\bibinfo {title} {Handbook of Mathematical
  Functions: With Formulas, Graphs, and Mathematical Tables}}}\ (\bibinfo
  {publisher} {Dover Publications},\ \bibinfo {year} {1964})\BibitemShut
  {NoStop}%
\end{thebibliography}%


\end{document}